\documentclass[lettersize,journal]{IEEEtran}
\usepackage{amsmath,amsfonts}

\usepackage{array}
\usepackage[caption=false,font=normalsize,labelfont=sf,textfont=sf]{subfig}
\usepackage{textcomp}
\usepackage{stfloats}
\usepackage{url}
\usepackage{verbatim}
\usepackage{graphicx}
\usepackage{hyperref}
\usepackage{hyperref}
\hypersetup{
    colorlinks=true,
    linkcolor=blue,
    filecolor=magenta,      
    urlcolor=blue,
    citecolor = blue,
}
\usepackage{cite}

\usepackage[normalem]{ulem}
\usepackage{multirow}
\usepackage{xcolor}
\usepackage{tcolorbox}
\definecolor{customgreen}{HTML}{D9FFD9}
\definecolor{customblue}{HTML}{D9D9FF}
\usepackage{bbding}
\usepackage{algorithm}
\usepackage{algpseudocode}
\usepackage{pifont}
\usepackage{ulem}

\newtheorem{theorem}{Theorem}
\makeatletter
\newcommand{\settheoremtag}[1]{
  \let\oldthetheorem\thetheorem
  \renewcommand{\thetheorem}{#1}
  \g@addto@macro\endtheorem{
    \addtocounter{theorem}{-1}
    \global\let\thetheorem\oldthetheorem}
  }
\makeatother

\newtheorem{remark}[theorem]{Remark}

\newcommand{\leons}[1]{{\color{magenta}{#1}}}

\def\BibTeX{{\rm B\kern-.05em{\sc i\kern-.025em b}\kern-.08em
    T\kern-.1667em\lower.7ex\hbox{E}\kern-.125emX}}

\usepackage{balance}

\let\titleold\title
\renewcommand{\title}[1]{\titleold{#1}\newcommand{\thetitle}{#1}}
\def\maketitlesupplementary
   {
   \newpage
       \twocolumn[
        \centering
        \huge
        Supplementary Material: \thetitle \\
        \vspace{0.5em}
        
        \vspace{1.0em}
       ] 
   }

\begin{document}
\title{Distilling Knowledge for Designing Computational Imaging Systems}


\author{Leon Suarez-Rodriguez, ~\IEEEmembership{Student Member,~IEEE}, Roman Jacome~\IEEEmembership{Student Member,~IEEE}, Henry Arguello,~\IEEEmembership{Senior Member,~IEEE}
\thanks{Leon Suarez-Rodriguez and Henry Arguello are with the Department of Systems Engineering and Informatics at the Universidad Industrial de Santander}
\thanks{Roman Jacome is with the Department of Electrical, Electronics and
Telecommunications Engineering at the Universidad Industrial de Santander}
\thanks{The conference precursor of this work was presented at the 2024 IEEE International Conference on Image Processing (ICIP) \cite{suarez2024highly}.}}



\maketitle

\begin{abstract}

Designing the physical encoder is crucial for accurate image reconstruction in computational imaging (CI) systems. Currently, these systems are designed via end-to-end (E2E) optimization, where the encoder is modeled as a neural network layer and is jointly optimized with the decoder. However, the performance of E2E optimization is significantly reduced by the physical constraints imposed on the encoder. Additionally, since the E2E learns the parameters of the encoder by backpropagating the reconstruction error, it does not promote optimal intermediate outputs and suffers from gradient vanishing. To address these limitations, we reinterpret the concept of knowledge distillation (KD) for designing a physically constrained CI system by transferring the knowledge of a pretrained, less-constrained CI system. Our approach involves three steps: (1) Given the original CI system (student), a teacher system is created by relaxing the constraints on the student's encoder. (2) The teacher is optimized to solve a less-constrained version of the student's problem. (3) The teacher guides the training of the student through two proposed knowledge transfer functions, targeting both the encoder and the decoder feature space. The proposed method can be employed to any imaging modality since the relaxation scheme and the loss functions can be adapted according to the physical acquisition and the employed decoder. This approach was validated on three representative CI modalities: magnetic resonance, single-pixel, and compressive spectral imaging. Simulations show that a teacher system with an encoder that has a structure similar to that of the student encoder provides effective guidance. Our approach achieves significantly improved reconstruction performance and encoder design, outperforming both E2E optimization and traditional non-data-driven encoder designs.

\end{abstract}


\section{Introduction}
\IEEEPARstart{C}{omputational} imaging (CI) systems have extended the capabilities of traditional imaging systems by integrating computational algorithms with physical image acquisition,   surpassing conventional imaging limitations such as blurring, dynamic range, spatial resolution and depth of field \cite{bhandari2022computational, coded_optical_imaging_book}. CI systems employ a physical encoder to encode the desired high-dimensional signal information into low-dimensional coded projections.  A computational decoder then decodes these projections to reconstruct the underlying high-dimensional signal \cite{E2E_PROF_HENRY}. CI systems have been applied in {several} imaging fields such as optical imaging, seismic imaging, computational microscopy,  and medical imaging \cite{bhandari2022computational, CI_SIGNAL_PROCESSING, COI_NATURE}. The performance of these systems relies on the effective design of the encoder, as it determines how the scene is sampled and encoded, affecting the quality and amount of information that can be encoded and subsequently reconstructed by the computational decoder.

{The traditional design of CI systems has relied on random or structured patterns, which have proven effective in exploiting wavelength and depth coding across various imaging modalities. For example, coded apertures (CAs) have used patterns like Hadamard and Fourier bases \cite{zhang2017hadamard}, and magnetic resonance imaging (MRI) has employed undersampling patterns such as radial and spiral \cite{CS_MRI, INTRO_COMPRESIVE_SAMPLING, SPC_COMPRESSIVE_SAMPLING}. Regular and jitter sampling patterns have been used for seismic survey acquisition \cite{maria_semismic}. In diffractive optical elements (DOEs), examples include Fresnel \cite{FRESNEL_LENS}, saddle \cite{SADDLE_LENS}, and spiral \cite{spiralphaseplate} lenses. Similarly, for 3D ultrasound imaging, 2D sparse arrays have been designed using deterministic or stochastic spiral distributions \cite{ramalli2022design}. Furthermore, analytical designs leveraging compressive sensing principles have been widely explored across various applications \cite{colored_CA, CASSI_PROFE_HENRY, INCOHERENCE_UNDERSAMPLING_MRI, POISSON_MRI, COMPRESSIVE_SEISMIC_SURVEY, COMPRESSIVE_SEISMIC_SURVEY_2}. While these traditional designs have been effective in their respective contexts, they are typically limited by predefined patterns, limiting their performance.}

In contrast, current approaches optimize CI systems in a data-driven manner using an end-to-end (E2E) framework, enabling high-performance task-specific CI systems. Here, the encoder and the computational decoder are jointly optimized for a specific imaging task. These methods leverage deep learning by representing the image formation model as a neural network layer, with subsequent layers serving as the computational decoder \cite{E2E_PROF_HENRY}. {Examples include the design of a DOE 
for high dynamic range imaging \cite{PSF_HDR}, the design of a colored CA for depth estimation \cite{JHON_DEPTH}, undersampling mask optimization for MRI reconstruction \cite{MRI_MASK_OPT}, seismic survey design for seismic imaging reconstruction \cite{maria_semismic}, and microscopic illumination pattern design 
for image reconstruction \cite{ILLUIMATION_PHASE_IMAGING}.}

The design of the encoder must account for physical constraints. In MRI, for instance, long scans can lead to patient discomfort, motion artifacts, high costs, and long waiting times that may extend for months \cite{MRI_MASK_OPT, MRI_EFFICIENT, MRI_MOTION}. To address these issues, $k-\text{space}$ measurements are undersampled to reduce acquisition time, resulting in a trade-off between obtaining high-quality images and minimizing scan times. 

Similarly, optical imaging systems employ CAs to reduce the amount of information required during acquisition while maintaining high-quality image reconstruction. {Binary-valued CAs are commonly used because they are easier to fabricate and calibrate than real-valued or quantized CAs, use less integration time, and have lower storage requirements  \cite{E2E_JORGE_IEEE}}. However, each element of the binary-valued CA is limited to capturing only two states: blocking or passing light. {Compared to other CA types, this limitation reduces} the amount of information that can be encoded and subsequently reconstructed, resulting in less detailed or lower-quality images \cite{E2E_PROF_HENRY, E2E_JORGE_IEEE}.
Furthermore, CA imaging systems are constrained by the need to acquire a limited number of snapshots to reduce acquisition and processing times, resulting in a trade-off between image quality and the time needed for acquisition and processing.  Therefore, these systems aim to minimize the number of acquisitions and maximize the performance on the imaging task \cite{colored_CA, E2E_JORGE_IEEE}. 

These limitations of CI systems are incorporated into the E2E optimization problem through regularization functions that restrict the set of optimal values that the parameters of the imaging formation model can take \cite{E2E_PROF_HENRY}. Consequently, these regularization functions reduce the degrees of freedom of the CI system during training, ultimately degrading its performance on the imaging task \cite{SPC_ROMAN}. Furthermore, since the E2E optimization framework learns the encoder parameters by backpropagating the reconstruction error, it does not promote optimal intermediate outputs in the computational decoder. Moreover, the encoder of the CI system may not be optimal as it suffers from the vanishing gradient problem due to being the first layer of the E2E framework \cite{SPC_ROMAN}. Thus, it is required to develop new learning techniques to address the implementation constraints of the encoder and find optimality criteria for the design of the encoder and the decoder's intermedium layers.

To address the performance limitations of E2E optimization for CI system design, this work proposes a novel approach inspired by {knowledge distillation (KD) \cite{KD_HINTON} theory}. In KD, a larger and more complex neural network, known as the teacher, guides the training of a smaller and simpler network, called the student. The student is specifically designed for resource-constrained environments, and with the teacher’s guidance, it can achieve higher performance than it would through standalone training. {We reinterpret the concept of KD for CI system design, where} a high-performance, less-constrained CI system serves as the teacher model, while a more physically constrained CI system acts as the student model. Both CI systems share computational decoders with the same neural network architecture, with the focus on addressing the constraints in the student’s encoder. For the simulations, we use the well-known U-Net architecture \cite{UNET}; however, the proposed methodology can be extended to any computational decoder neural network architecture. The proposed approach follows a three-stage methodology, {illustrated in Figure \ref{fig:KD_SCHEME}}: (1) the teacher encoder is created by relaxing the constraints on the student encoder, (2) the teacher encoder is optimized along with its reconstruction decoder to solve a less-constrained version of the student’s problem, and (3) the teacher provides guidance to the student by transferring knowledge through various proposed KD loss functions. 

{The proposed methodology is designed to be general-purpose for CI system design and serves as an alternative to E2E optimization. In this approach the student is designed to be implementable for real-world acquisition, addressing practical physical constraints in the encoder. However, the teacher system is not required to be practical for real-world acquisition, as it is only used in simulation to transfer its knowledge to the student. This approach allows for endless possibilities in creating the teacher, such as relaxations in the number of measurements, codification in non-conventional bases (e.g., complex or quaternion), or even using a different CI system as the teacher to distill the student, offering more flexibility than E2E optimization.} 
Preliminary results for this method were presented in \cite{suarez2024highly} for single-pixel imaging. 

\leons{\begin{figure}[t!]
    \centering
    \includegraphics[width=\columnwidth]{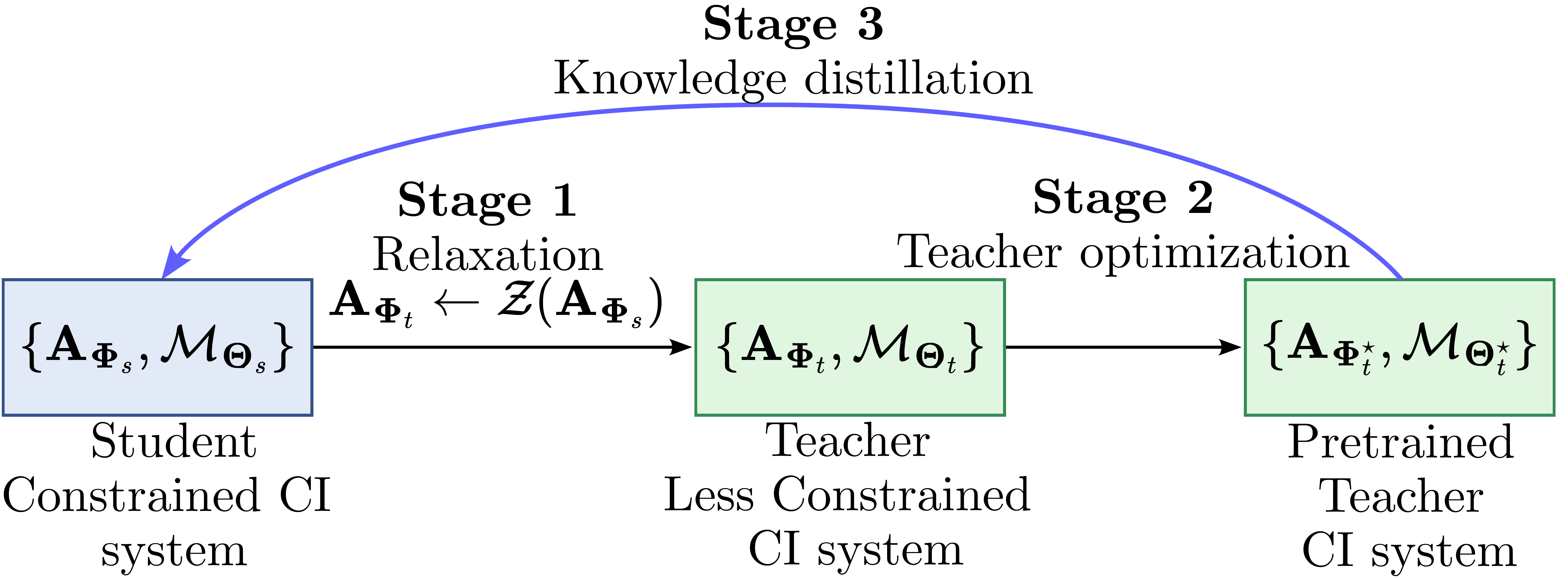}
    \caption{The student system is a constrained CI system with encoder $\mathbf{A}_{\boldsymbol{\Phi}_s}$ and decoder $\mathcal{M}_{\boldsymbol{\Theta}_s}$. In the first stage, by relaxing the student encoder constraints, a teacher encoder $\mathbf{A}_{\boldsymbol{\Phi}_t}$ is derived. In the second stage, the teacher encoder and its reconstruction network $\mathcal{M}_{\boldsymbol{\Theta}_t}$ are optimized to solve a less-constrained version of the student's problem, resulting in ${\mathbf{A}_{\boldsymbol{\Phi}_t^\star}, \mathcal{M}_{\boldsymbol{\Theta}_t^\star}}$. In the third stage, the knowledge of the pretrained teacher system is used to guide and enhance the performance of the student system’s encoder and decoder.}
    \label{fig:KD_SCHEME}
\end{figure}}

{We validate {the} effectiveness} {of the proposed approach} on three representative CI systems: MRI, the single-pixel camera (SPC) \cite{SPC_COMPRESSIVE_SAMPLING}, and the single disperser coded aperture snapshot spectral imager (SD-CASSI) \cite{SDCASSI}. 
We devise different design criteria for relaxing the {student's {encoder} to obtain the} teacher's {encoder} {for} each system. In MRI we created a teacher with a lower acceleration factor ($AF$) than the student. {For} the SPC {and SD-CASSI systems,} the teacher uses real-valued CAs and a higher number of snapshots than the binary-valued CA student. However, the design of the teacher is still an open research question, where different encoder parametrizations can be useful for knowledge transfer. 

\begin{figure*}[t!]
    \centering
    \includegraphics[width=0.95\textwidth]{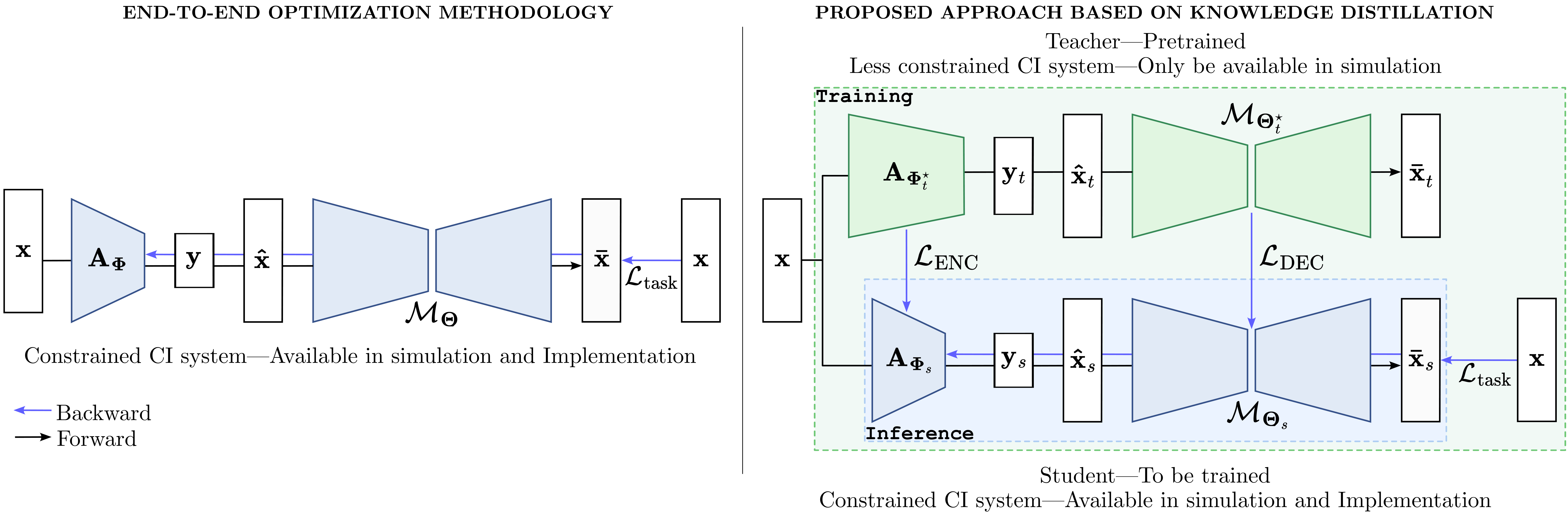}
    \caption{Comparative of E2E optimization and the Proposed KD methodology for designing CI systems. During training, the pretrained teacher guides the learning of the student through the proposed loss functions $\mathcal{L}_{\text{ENC}}$ and $\mathcal{L}_{\text{DEC}}$. For inference, the student system operates independently.}
    \label{fig:KDvsE2E}
\end{figure*}

Two types of loss functions were employed to transfer knowledge from the teacher system to the student system{, as shown in Figure \ref{fig:KDvsE2E}: the encoder and decoder loss functions.} The encoder loss function aligns the student's encoder structure with the teacher's encoder structure, while the decoder loss function aligns the feature space of the student's decoder with the teacher's. The encoder loss function mitigates the vanishing gradient problem in the encoder, as this loss only affects the {encoder}. The decoder loss guides the learning of optimal intermediate features in the student’s computational decoder. {This is achieved because the teacher's encoder operates under fewer constraints, enabling higher recovery performance in the teacher's decoder compared to that of the student. As a result, the teacher's decoder produces more robust learned feature representations, which can effectively guide the student's feature space}. Simulation results indicate that the nature of the teacher system's encoder significantly impacts the performance of the student system.  Specifically, a teacher with a {relaxation close to that of} the student {(similar number of measurements and similar nature of the codification basis)} provides effective guidance, leading to better {encoder} design and reconstruction performance than E2E optimization. In MRI, when used as fixed {the encoder} to train a reconstruction neural network, the student's learned undersampling mask achieved better reconstruction performance. In the SPC, the {sensing matrix} demonstrated lower condition numbers and mutual coherence. In the SD-CASSI, improved spectral band correlation was observed.

To summarize, our contributions are the following:
\begin{itemize} 
\item We introduce a novel KD framework for CI system design, addressing the performance limitations imposed by physical constraints on the {physical encoder} in traditional E2E optimization. In this approach, the constraints on the {encoder} in the highly constrained student are relaxed to obtain the teacher, which is then trained and used to guide the learning of the student. 
\item  The knowledge is transferred from the teacher to the student to two proposed loss functions: the encoder loss, which aligns the {encoder} structure of the student with that of the teacher; and the decoder loss, which aligns the feature spaces of their respective computational decoders.
\item We demonstrate that teachers with a similar number of measurements and with {an encoder} of a similar nature {of the codification basis} to those of the student provide effective guidance, leading to improved performance in both the design of the student's {encoder} and its reconstruction capabilities.
\item We validate the proposed KD approach across three widely used CI systems—MRI, SPC, and SD-CASSI—showing superior performance compared to traditional E2E optimization, and demonstrating that the proposed methodology can be easily adapted to the design of any CI system. \end{itemize}

\section{Background}

\subsection{End-to-End Optimization}

CI systems can be interpreted as an encoder-decoder framework, where the imaging formation model encodes a scene $\mathbf{x} \in \mathbb{R}^n$ into coded projections $\mathbf{y} \in \mathbb{R}^m$, with $m \ll n$ \cite{COI_NATURE}. This process follows the forward sensing model: 

\begin{equation}
    \label{eq:forward}
    \mathbf{y}=\mathbf{A}_{\boldsymbol{\Phi}} \mathbf{x} + \boldsymbol{\eta},
\end{equation}

\noindent where $\mathbf{A}_{\boldsymbol{\Phi}} \in \mathbb{R}^{m\times n}$ represents the {encoder} forward model of the CI system, parametrized by $\boldsymbol{\Phi}$, and $\boldsymbol{\eta} \in \mathbb{R}^m$ is additive noise. A computational decoder then recovers the underlying signal $\mathbf{x}$. The state-of-the-art approach to optimize CI systems involves an E2E deep learning optimization, where the physics-based forward imaging model is incorporated into deep learning architectures as a differentiable layer \cite{E2E_PROF_HENRY, COI_NATURE}. By modeling both the sensing model $\mathbf{A}_{\boldsymbol{\Phi}}$ parametrized by $\boldsymbol{\Phi}$ and the computational decoder $\mathcal{M}_{\boldsymbol{\Theta}}$ parametrized by $\boldsymbol{\Theta}$ as neural network layers, the E2E optimization problem is formulated as follows:


\begin{equation}
\label{eq:e2e_optimization}
\begin{aligned}
 \{\boldsymbol{\Theta^\star}, \boldsymbol{\Phi^\star}\} 
     =&\underset{\substack{\boldsymbol{\Theta}, \boldsymbol{\Phi} }}{\arg \min } \ \frac{1}{P} \sum_{p=1}^P  \| \mathcal{M}_{\boldsymbol{\Theta}}(\mathbf{A}_{\boldsymbol{\Phi}}^\top \mathbf{A}_{\boldsymbol{\Phi}} \mathbf{x}_p) -\mathbf{x}_p \|_2^2 \\&+ \tau \mathcal{R}_{\tau}(\mathbf{\Phi}) + \mu \mathcal{R}_{\mu}(\boldsymbol{\Theta}),\\
\end{aligned}
\end{equation}

\noindent where $\{\boldsymbol{\Theta^\star}, \boldsymbol{\Phi^\star}\}$ are the set of optimal values of the encoder and the computational decoder parameters, respectively, $\{\mathbf{x}_p\}_{p=1}^P$ is the dataset, and $\mathcal{R}_{\tau}(\mathbf{\Phi})$ is a regularization function that acts on the parameters of the imaging formation model to promote specific physical constraints, such as binarization and quantization for CAs {in systems like the SPC and SD-CASSI}, transmittance for the  $k-\text{space}$ undersampling mask with a given acceleration factor in MRI, the number of angles in computed tomography, or the quantization of the height map in DOEs. The regularization parameter $\tau$ balances the trade-off between the desired constraints promoted by the regularization function and the performance of the recovery \cite{E2E_PROF_HENRY}. Additionally, the regularization function $\mathcal{R}_{\mu}(\boldsymbol{\Theta})$ prevents overfitting of the computational decoder parameters $\boldsymbol{\Theta}$ to the training data {by limiting the network's complexity through penalizing large weight values}, where $\mu$ as its regularization parameter. 

The regularization functions $\mathcal{R}_{\tau}(\mathbf{\Phi})$ used to enforce the physical constraints of the CI systems in the E2E optimization restrict the set of optimal values that the imaging formation model can take. This reduces the degrees of freedom during training, ultimately limiting the recovery performance of the CI system. This performance reduction occurs due to the influence of these physical constraints on the structure of the imaging formation model, which affects the effectiveness of the computational decoder \cite{E2E_JORGE_IEEE, SPC_ROMAN}. {Furthermore, the encoder is susceptible to vanishing gradients as it is the first layer in the E2E framework, which further limits its ability to learn an optimal codification}. Therefore, effective design of the {encoder} is essential for achieving high-performance recovery in CI systems.

Equation \eqref{eq:e2e_optimization} is solved using gradient descent-based algorithms, and once the optimal encoder values are learned $\boldsymbol{\Phi^\star}$, the physical imaging formation model can be implemented to acquire real measurements. Subsequently, the learned computational decoder $\mathcal{M}_{\boldsymbol{\Theta}^\star}$ is used to perform the recovery task on the acquired measurements \cite{E2E_PROF_HENRY,COI_NATURE}.

\subsection{Knowledge Distillation}

{KD, popularized by \cite{KD_HINTON},  is an effective technique for optimizing neural networks for inference on resource-constrained devices. In KD a smaller and simpler model, known as the student, is trained to mimic the behavior of a larger and more complex model, known as the teacher, leading to the student model achieving better performance than would be possible if it were trained directly on the task \cite{KD_REVIEW, KD_INTERMEDIATE_FEATURES}. KD is especially valuable in scenarios where the deployment of neural network models for real-time inference is challenging due to physical constraints or system limitations, such as in resource-constrained devices like smartphones, smartwatches, or embedded devices \cite{ci_review}, where storage and computational resources are limited. }

{Knowledge transfer in KD can occur through various sources, including intermediate feature maps of the teacher model \cite{KD_INTERMEDIATE_FEATURES}, relationships between consecutive feature maps \cite{KD_GRAM_MATRIX}, and the response of the teacher model \cite{KD_HINTON}. Feature maps represent outputs from intermediate layers, while the response refers to the output of the model's final layer \cite{KD_APPLICATIONS}. The KD optimization problem can be expressed as:}

{\begin{align}
    \boldsymbol{\Theta}_s^\star =&\underset{\substack{\boldsymbol{\Theta}_s}}{\arg \min } \frac{1}{P}\sum_{p=1}^P \lambda_1 \mathcal{L}_{\text{task}}(\mathcal{M}_{\boldsymbol{\Theta}_s}(\mathbf{x}_p), \mathbf{d}_p) \nonumber\\&+ \lambda_2 \mathcal{L}_{\text{KD}}(\mathcal{M}_{\boldsymbol{\Theta}_s}(\mathbf{x}_p), \mathcal{M}_{\boldsymbol{\Theta}_t^\star}(\mathbf{x}_p))
\end{align}}

{\noindent where $\boldsymbol{\Theta}_s$ represents the parameters of the student model, $\boldsymbol{\Theta}_t^\star$, denotes the optimized parameters of the teacher model, $\mathbf{x}_p$ and $\mathbf{d}_p$ are the input and desired output of the model, $\mathcal{L}_{\text{task}}$ is the task-specific loss (e.g., reconstruction or classification), and $\mathcal{L}_{\text{KD}}$ is the KD loss that transfers knowledge from the teacher to the student. The regularization parameters $\lambda_1$ and $\lambda_2$ balance the contributions of the task-specific and KD losses.}

Although KD has been applied mainly to high-level computer vision tasks such as segmentation, detection, and classification, its applications in low-level vision tasks such as reconstruction and denoising have been less explored \cite{DISTILLATION_SUPER_RES, DISTILLATION_SUPER_RES_2, DISTILLATION_SUPER_RES_3, DISTILLATION_SUPER_RES_4, DISTILLATION_SUPER_RES_5, KD_DENOISING}. Furthermore, the use of KD in CI tasks has received limited attention. For example,  \cite{MRI_KD_FIXED_MASKS} applied KD to reconstruct magnetic resonance images but did not address the design of the undersampling mask. Similarly, \cite{PHASE_RETR_KD} utilized KD to enhance the performance of a phase retrieval system's recovery network without considering the design of the codification masks, while \cite{video_snapcap} employed KD for compressive video captioning without optimizing the CAs. However, to the best of our knowledge, KD has not been explored to tackle challenges in the encoder design.

\section{Acquisition models}

The proposed approach has been validated in three popular CI systems, MRI, the SPC, and the SD-CASSI. The following subsections provide descriptions of these systems.

\subsection{Magnetic resonance imaging}

MRI is a non-invasive imaging technique that captures images of the body using strong magnetic fields and radio waves. Unlike X-ray imaging, it does not employ harmful ionizing radiation \cite{CI_SIGNAL_PROCESSING}.  MRI scans can take a long time depending on the part of the body being imaged, which can cause discomfort to the patient, increase costs, and cause motion artifacts. Additionally, this extended scan time limits the number of patients that can be imaged in a day, resulting in long waiting times that can stretch for weeks or even months \cite{MRI_MASK_OPT,MRI_EFFICIENT}. Therefore, the acquisition process is undersampled to reduce scan times, and magnetic resonance images are reconstructed from the undersampled $k-\text{space}$ measurements. Traditional undersampling approaches use fixed undersampling patterns, with the most common including random rectilinear, equispaced rectilinear, cartesian, radial, and spiral \cite{MRI_SAMPLING_SCHEMES}. However, each undersampling pattern may lead to different reconstruction performances for the same reconstruction method \cite{MRI_MASK_OPT, CS_MRI, MRI_SAMPLING_SCHEMES}. More recent approaches optimize the undersampling masks pattern together with the reconstruction neural network in an E2E manner \cite{MRI_MASK_OPT, weber2024constrained, E2E_MRI_2, E2E_MRI_3}. 
The single-coil MRI acquisition process is given by equation \eqref{eq:forward}, where $\mathbf{A}_{\boldsymbol{\Phi}}=\boldsymbol{\Phi}\mathbf{F}$, where $\boldsymbol{\Phi} \in \{0,1\}^{m\times n}$ denotes an undersampling mask  and $\mathbf{F}\in \mathbb{C}^{n\times n}$ is the Fourier transform operator. In this work, the $k-\text{space}$ undersampling mask is modeled as a neural network layer with trainable parameters {$\mathbf{W} \in \mathbb{R}^{m\times n}$}, incorporating a Heaviside step activation function $H(x) = \frac{x+|x|}{2|x|}$  to binarize the mask, such that {$\boldsymbol{\Phi}=H(\mathbf{W})\in \{0,1\}^{m\times n}$}. However, this activation function is not differentiable at zero and has a gradient of zero elsewhere, making it unsuitable for learning the undersampling mask directly. To address this, a Straight-Through Estimator (STE) is employed \cite{STE_gradient}. The STE enables the gradient from the preceding layer to propagate through the activation function, facilitating the optimization of the binary undersampling mask, such that $\frac{\partial \boldsymbol{\Phi}}{\partial\mathbf{W}}=\mathbf{I}$.

To promote the desired acceleration factor for the $k-\text{space}$ undersampling mask $\boldsymbol{\Phi}$ the following regularization term is employed in \eqref{eq:e2e_optimization} as:

\begin{equation}
    \mathcal{R}_{\tau}(\boldsymbol{\Phi}) = \tau \left( \frac{\Vert\boldsymbol{\Phi}\Vert_0}{n} - \frac{1}{AF} \right)^4.
    \label{eq:reg_tramittance}
\end{equation}

\subsection{Single pixel camera}
The SPC consists of an objective lens that forms an image of the scene onto a CA, which spatially modulates the scene. The encoded image is then focused by a collimator lens onto a single-photon detector. To obtain different measurements (snapshots) from the same scene, the CA is changed for each snapshot \cite{SPC_JORGE}.


The forward model of the SPC is given by equation \eqref{eq:forward}. Here, the forward operator matrix is built upon the vectorization of the CAs ${\boldsymbol{\Phi}}_i$ as $\mathbf{A}_{\boldsymbol{\Phi}} = [{\boldsymbol{\Phi}}_1, \dots, {\boldsymbol{\Phi}}_m]^\top$. The relationship between the number of acquired measurements and the image size is given by the compression ratio $\gamma = \frac{m}{n}$.

The CAs can be binary-valued or real-valued. In a binary-valued CA, translucent or opaque elements are used to either block or allow light to pass through. Each entry of the CA takes a value of $0$ or $1$ ($\boldsymbol{\Phi}_{i,j} \in \{0,1\}$), representing the blocking or passage of light \cite{E2E_JORGE_OPTICA, CAS_PROF_HENRY}. On the other hand, real-valued CAs attenuate light at different levels, admitting a wider range of values compared to binary CAs. In a real-valued CA, each entry takes a value $\boldsymbol{\Phi}_{i,j} \in \mathbb{R}$ \cite{E2E_JORGE_IEEE}.  
The SPC system aims to minimize the number of snapshots, reducing acquisition and processing times while maximizing reconstruction performance.

In this work, the CAs of the SPC were modeled as neural network layers with trainable parameters {$\mathbf{W}_i \in \mathbb{R}^{n}, i=1,\dots,m$}. The binary-valued CAs are obtained as {$\boldsymbol{\Phi}_i = \operatorname{sign}(\mathbf{W}_i) \in \{-1,1\}^n, i=1,\dots,m$} where $\operatorname{sign}(\cdot)$ denotes the element-wise sign function $\operatorname{sign}(x) = \frac{x}{|x|}$. However, this function, like the Heaviside step, is not differentiable at zero and has a gradient of zero elsewhere, leading to the same optimization issues. Therefore, we use an STE, such that $\frac{\partial \boldsymbol{\Phi}}{\partial\mathbf{W}}=\mathbf{I}$.

\subsection{Single disperser Coded aperture snapshot spectral imager}

The SD-CASSI system \cite{SDCASSI} is a snapshot spectral imaging system where the incoming light is modulated spatially using a CA. Subsequently, a disperser element, such as a prism or a grating, disperses the spectral information of each spatial location across a wide area on the detector, resulting in the multiplexation of the spatial and spectral information of the scene.

The forward model of the SD-CASSI is expressed by equation \eqref{eq:forward}. Here  $\mathbf{x} \in \mathbb{R}^{N M  L}$ represents the vectorized data cube with spatial dimensions $N \times M$ and $L$ spectral bands with $n=NML$, $\mathbf{y} \in \mathbb{R}^{N(M+L-1)}$ denotes the vectorized detector measurement, and $\boldsymbol{\eta} \in \mathbb{R}^{N(M+L-1)}$ is additive noise. 
The sensing matrix that accounts for the coding and dispersing effect is given by  $\mathbf{A}_{\boldsymbol{\Phi}}=\mathbf{P}\boldsymbol{\Phi} \in \mathbb{R}^{N(M+L-1)\times NML}$, where $\mathbf{P} \in \mathbb{R}^{N(M+L-1)\times NML}$  is a matrix that models the dispersion and $\boldsymbol{\Phi} \in \mathbb{R}^{NML \times NML}$ is the diagonalized CA. To acquire a set of $m$ snapshots $\mathbf{y}=[(\mathbf{y}^0)^\top, \dots, (\mathbf{y}^{m-1})^\top]^\top$ represents the set of measurements,  $\mathbf{A}_{\boldsymbol{\Phi}}=[(\mathbf{A}_{\boldsymbol{\Phi}}^0)^\top, \dots ,(\mathbf{A}_{\boldsymbol{\Phi}}^{m-1})^\top]^\top$ are the set of sensing matrices, and $\boldsymbol{\boldsymbol{\eta}}=[(\boldsymbol{\eta}^0)^\top, \dots, (\boldsymbol{\eta}^{m-1})^\top]^\top$ is the set of noise vectors. The SD-CASSI shares similar constraints with the SPC system. In the SD-CASSI, CAs may be binary-valued ($\boldsymbol{\Phi}_{i,j} \in \{0, 1\}$) or real-valued ($\boldsymbol{\Phi}_{i,j} \in [0,1]$), additionally, the number of snapshots is minimized to reduce acquisition and processing times while maximizing recovery performance. To model the CAs for the SD-CASSI in the E2E framework, the same process done for the SPC is applied using the Heavyside step activation function $H$ along with a STE.


\section{Distilling knowledge for CI system design}

To address the performance limitations of CI systems optimized through E2E optimization, a KD approach is proposed. In this approach, a less constrained CI system guides the learning of a more constrained CI system through knowledge transfer. Specifically, the proposed approach is divided into three stages. First, the teacher system encoder is obtained as a relaxation of the student system encoder, such that $\mathbf{A}_{\boldsymbol{\Phi}_t} \gets \mathcal{Z}(\mathbf{A}_{\boldsymbol{\Phi}_s})$, where $\mathcal{Z}(\cdot)$ is a relaxation operator. Second, the teacher is optimized following Eq. \eqref{eq:e2e_optimization}, solving a relaxed version of the student's recovery problem, obtaining the optimal parameters of the encoder and decoder $\{\boldsymbol{\Theta}_t^\star, \boldsymbol{\Phi}_t^\star\}$. Third, once trained, the teacher guides the student's learning process through proposed knowledge transfer functions. The proposed KD optimization pipeline is illustrated in algorithm \ref{alg:KD_pipeline}.

Before describing the proposed KD optimization scheme, the nature of the teacher encoder is first defined, and its key differences from the student encoder are outlined. Next, we describe the employed computational decoder.

\begin{algorithm}[!t]
\caption{Proposed KD optimization of CI systems}\label{alg:KD_pipeline}
\begin{algorithmic}[1]
\small
\Require $\mathbf{A}_{\boldsymbol{\Phi}_s}, \mathcal{M}_{\boldsymbol{\Theta}_s}, N$  \Comment{Student's encoder and computational decoder, number of epochs}

\State \textbf{Stage 1:} Obtain the teacher's encoder $\mathbf{A}_{\boldsymbol{\Phi}_t}$ by relaxing the student's encoder constraints with operator $\mathcal{Z}(\cdot)$:
\State $\mathbf{A}_{\boldsymbol{\Phi}_t} \gets \mathcal{Z}(\mathbf{A}_{\boldsymbol{\Phi}_s})$  

\Statex

\State \textbf{Stage 2:} Optimize the teacher's encoder and decoder parameters

\For {$i$ in $1,2,\dots, N$}

\State $\mathcal{L}_{\text{task}}=\left\|\mathcal{M}_{\boldsymbol{\Theta}_t^i} \left(\mathbf{A}_{\boldsymbol{\Phi}_t^i}^\top\mathbf{A}_{\boldsymbol{\Phi}_t^i} \mathbf{x}\right)- \mathbf{x}\right\|_2^2$ 

\State $\mathcal{L} = \mathcal{L}_{\text{task}} + \tau \mathcal{R}_{\tau}(\boldsymbol{\Phi}_t^i) + \mu\mathcal{R}_{\mu}(\boldsymbol{\Theta}_t^i)$

\State \small $\boldsymbol{\Theta}_t^{i+1} = \boldsymbol{\Theta}_t^{i} - \alpha \left( \frac{\partial \mathcal{L}_{\text{task}}}{\partial \boldsymbol{\Theta}_t^i} + \mu \frac{\partial \mathcal{R}_{\mu}(\boldsymbol{\Theta}_t^i)}{\partial \boldsymbol{\Theta}_t^i} \right)$  

\State \small $\boldsymbol{\Phi}_t^{i+1} = \boldsymbol{\Phi}_t^{i} - \alpha \left( \frac{\partial\mathcal{L}_{\text{task}}}{\partial \boldsymbol{\Phi}_t^i} + \tau \frac{\partial \mathcal{R}_{\tau}(\boldsymbol{\Phi}_t^i)}{\partial \boldsymbol{\Phi}_t^i} \right)$  

\EndFor
\Statex
\State The teacher's encoder and decoder parameters $\{\boldsymbol{\Theta}_t^\star, \boldsymbol{\Phi}_t^\star\}$ are set to be frozen.
\Statex
\State \textbf{Stage 3:} Optimize $\mathbf{A}_{\boldsymbol{\Phi}_s}, \mathcal{M}_{\boldsymbol{\Theta}_s}$ with the guidance of $\mathbf{A}_{\boldsymbol{\Phi}_t^\star}, \mathcal{M}_{\boldsymbol{\Theta}_t^\star}$ using knowledge transfer functions $\mathcal{L}_{\text{ENC}}$ and $\mathcal{L}_{\text{DEC}}$.

\For {$i$ in $1,2,\dots, N$}

\State $\mathbf{S} = \mathcal{M}^{[l]}_{\boldsymbol{\Theta}_s}(\mathbf{A}_{\boldsymbol{\Phi}_s^i}^\top \mathbf{A}_{\boldsymbol{\Phi}_s^i} \mathbf{x})$  \Comment{Student's features at layer $l$}

\State $\mathbf{T} = \mathcal{M}^{[l]}_{\boldsymbol{\Theta}_t^\star}(\mathbf{A}_{{\boldsymbol{\Phi}_t^i}^\star }^\top \mathbf{A}_{{\boldsymbol{\Phi}_t^i}^\star} \mathbf{x})$  \Comment{Teacher's features at layer $l$}

\State $\mathcal{L}_{\text{ENC}}=\mathcal{D}\left(\mathbf{A}_{\boldsymbol{\Phi}_s^i}, \mathbf{A}_{{\boldsymbol{\Phi}_t^i}^\star},\mathbf{x}\right)$ \Comment{$\mathcal{D}$ is a distance function}

\State $\mathcal{L}_{\text{DEC}}=\| \mathbf{S} - \mathbf{T} \|_2^2$

\State $\mathcal{L}_{\text{task}}=\left\|\mathcal{M}_{\boldsymbol{\Theta}_s^i} \left(\mathbf{A}_{\boldsymbol{\Phi}_s^i}^\top \mathbf{A}_{\boldsymbol{\Phi}_s^i} \mathbf{x}\right)- \mathbf{x}\right\|_2^2$ 

\State $\mathcal{L}_{\text{KD}} = \lambda_1 \mathcal{L}_{\text{task}} + \lambda_2 \mathcal{L}_{\text{DEC}} + \lambda_3 \mathcal{L}_{\text{ENC}} + \tau \mathcal{R}_{\tau}(\boldsymbol{\Phi}_s^i) + \mu\mathcal{R}_{\mu}(\boldsymbol{\Theta}_s^i)$

\State \small $\boldsymbol{\Theta}_s^{i+1} = \boldsymbol{\Theta}_s^{i} - \alpha \left(\lambda_1  \frac{\partial \mathcal{L}_{\text{task}}}{\partial \boldsymbol{\Theta}_s^i} +\lambda_2 \frac{\partial \mathcal{L}_{\text{DEC}}}{\partial \boldsymbol{\Theta}_s^i} + \mu \frac{\partial \mathcal{R}_{\mu}(\boldsymbol{\Theta}_s^i)}{\partial \boldsymbol{\Theta}_s^i}  \right)$

\State \small $\boldsymbol{\Phi}_s^{i+1} = \boldsymbol{\Phi}_s^{i} - \alpha \left(\lambda_1 \frac{\partial\mathcal{L}_{\text{task}}}{\partial \boldsymbol{\Phi}_s^i} + \lambda_2 \frac{\partial\mathcal{L}_{\text{DEC}}}{\partial \boldsymbol{\Phi}_s^i}\right)$ 
\Statex \hspace{4em}$+ \alpha \left(\lambda_3 \frac{\partial\mathcal{L}_{\text{ENC}}}{\partial \boldsymbol{\Phi}_s^i}  + \tau \frac{\partial \mathcal{R}_{\tau}(\boldsymbol{\Phi}_s^i)}{\partial \boldsymbol{\Phi}_s^i} \right)$ 

\EndFor

\Statex

\Return $\{\mathbf{A}_{\boldsymbol{\Phi}_s^\star}, \mathcal{M}_{\boldsymbol{\Theta}_s^\star}\}$

\end{algorithmic}
\end{algorithm}

\subsection{Computational encoder}

The design of the encoder of a CI system is the most important part of the system, as it determines how the scene is sampled and encoded. This directly affects the amount and quality of the acquired information, which is critical for the performance of the computational decoder. We now describe the teacher and student encoder configurations.

\subsubsection{Teacher Encoder Nature}

A key aspect of the proposed approach is the teacher configuration. The teacher's encoder is obtained by relaxing the student's encoder constraints using the operator $\mathcal{Z}(\cdot)$, such that $\mathbf{A}_{\boldsymbol{\Phi}_t} \gets \mathcal{Z}(\mathbf{A}_{\boldsymbol{\Phi}_s})$. Furthermore, the teacher is a synthetic CI system used only in simulations to guide the student, and therefore, it does not need to be feasible for real-world acquisition. This flexibility enables the exploration of various teacher configurations based on the application. As a result, the teacher's encoder codifies richer information than the student's, enabling its computational decoder to achieve higher performance. Below, we mention the explored teacher encoder configurations:

\begin{itemize}
    \item \textbf{MRI:} The teacher system is obtained by relaxing the student's acceleration factor, i.e, $\boldsymbol{\Phi}_t \in \{0,1\}^{m_t\times n}, \boldsymbol{\Phi}_s \in \{0,1\}^{m_s\times n}$, with {$\frac{n}{m_s} > \frac{n}{m_t}$.}
    
    \item \textbf{SPC:} {The teacher system is obtained by relaxing the student's constraint on the CAs from binary-valued to real-valued ($[\boldsymbol{\Phi}_t]_{i,j} \in \mathbb{R}$), and increasing the amount of acquired information, such that $m_t \ge m_s$.}

    \item \textbf{SD-CASSI:} {The teacher systems follow the same relaxations explored for the SPC.}
\end{itemize}

\begin{remark}
Here we devised a set of design criteria for the teacher model in the employed CI systems, however, this is still an open research question. We devise some insights that can be valuable for other imaging modalities. For computed tomography, where we are required to design the source angles to be minimal to reduce the user's radiation exposure \cite{mao2018fast} or in scenarios of limited angle where only a small portion of the scene can be scanned \cite{liu2023dolce}, a teacher can employ a huge number of equispaced source angles or a higher resolution sensor. In diffractive optical imaging, the DOEs have quantized heights \cite{li2022quantization} or in some scenarios DOEs are implemented in deformable mirrors which have a very limited parametrization based on few Zernike polynomials \cite{urrea2024dodo} which reduces the degree of freedom in the optimization. In this scenario, the teacher could be a quantization-free DOE or employ a high number of Zernike polynomials. In compressive seismic imaging, one is required to reduce the number of sources to mitigate environmental effects and acquisition costs and time \cite{maria_semismic}. Thus, a teacher design can be a system that uses a high number of uniformly distributed sources.
\end{remark}

\subsubsection{Student Encoder Nature}

The student CI system encoder, designed for both simulation and real-world implementation, must account for practical constraints. In MRI, we considered students with undersampled $k-\text{space}$ masks corresponding to acceleration factors $AF_s \in \{4, 8, 16\}$. For SPC, the following compression ratios were considered, with the student using binary-valued CAs: $\gamma_s \in \{0.05, 0.1, 0.2, 0.3, 0.4\}$. For the SD-CASSI system, the focus is on a student with a single snapshot and a binary-valued CA.

\subsection{Computational decoder}

\begin{figure}[t!]
    \centering
    \includegraphics[width=\columnwidth]{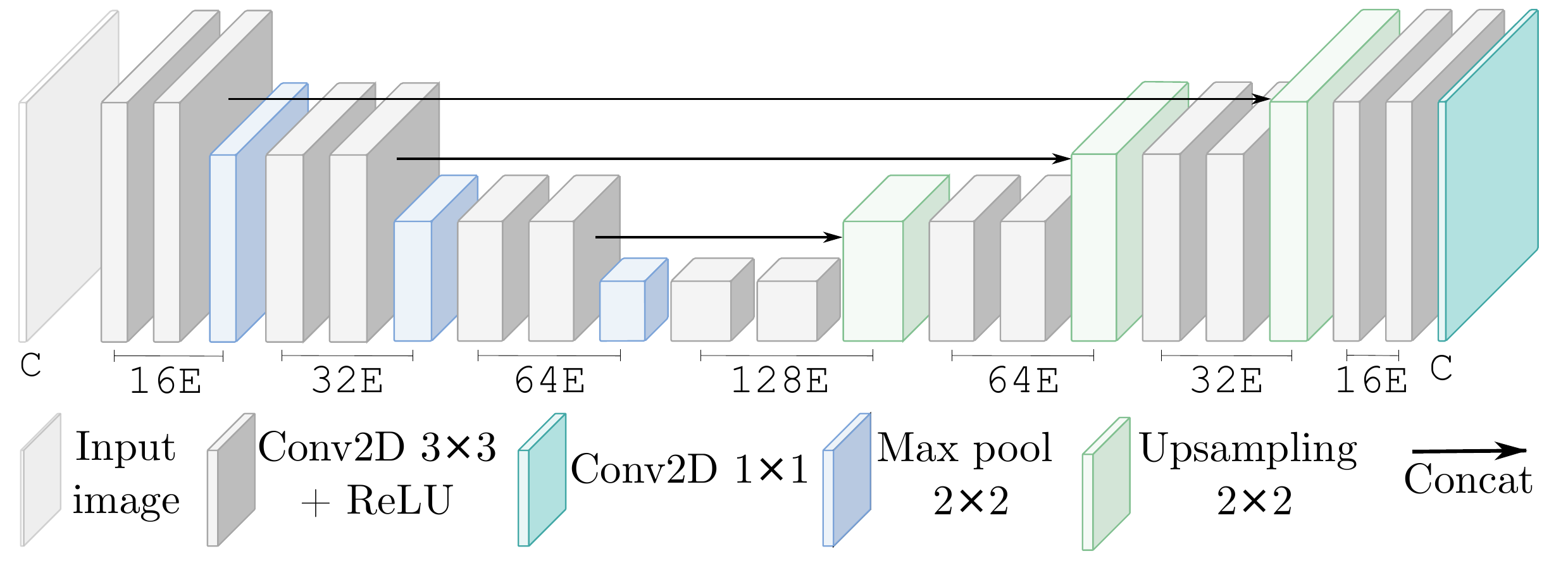}
    \caption{Scheme of the U-Net used as computational decoder, \texttt{C} is the number of channels of the input image, \texttt{C=2} for MR images, \texttt{C=1} for grayscale images, and \texttt{C=8} for the multi-spectral images.  \texttt{E} determines the number of filters of each convolutional layer of the U-Net, \texttt{E=1} for MRI,   \texttt{E=4} for the SPC, and \texttt{E=2} for the SD-CASSI system.}
    \label{fig:unet}
\end{figure}

Since we focus on developing a new methodology for designing CI systems rather than creating a novel signal reconstruction architecture, we used the well-known U-Net as the computational decoder for the experimental simulations. However, we emphasize that the proposed methodology can be extended to any computational decoder. Additionally, as our focus is on addressing the physical constraints in the {encoder} of the student, the computational decoders of both the student and teacher share the same U-Net architecture. {The U-Net architecture used in these simulations is depicted in Figure \ref{fig:unet}.}


\subsection{Knowledge Transfer Functions}
The transfer of knowledge from the teacher to the student is illustrated in Figure \ref{fig:KDvsE2E}. In this approach, the pretrained teacher system guides the student’s learning through two proposed types of loss functions: the encoder loss function $\mathcal{L}_{\text{ENC}}$ and the decoder loss function $\mathcal{L}_{\text{DEC}}$. Specifically, $\mathcal{L}_{\text{DEC}}$ promotes the alignment of the intermediate feature spaces between the teacher and student decoders, while $\mathcal{L}_{\text{ENC}}$ enforces alignment between the teacher and student encoders. These two functions allow guidance in both the sensing architecture and the computational recovery method, mitigating the vanishing gradient problem in the encoder and facilitating the learning of optimal intermediate outputs in the computational decoder. The learning of the student system involves minimizing the following loss function:

\begin{equation}
    \label{eq:KD_LOSS_TERMS}
    \mathcal{L}_{\text{KD}} = \lambda_1 \| \mathcal{M}_{\boldsymbol{\Theta}_s}(\mathbf{A}_{\boldsymbol{\Phi}_s}^\top \mathbf{A}_{\boldsymbol{\Phi}_s} \mathbf{x}) -\mathbf{x} \|_2^2  +  \lambda_2 \mathcal{L}_{\text{DEC}} + \lambda_3 \mathcal{L}_{\text{ENC}},
\end{equation}

\noindent where $\lambda_3 = 1 - \lambda_1 - \lambda_2$ with $\lambda_3 > 0$. The parameters $\lambda_1$, $\lambda_2$, and $\lambda_3$ are regularization terms that control the relative contributions of the different loss functions during optimization. Then, the KD optimization problem is then formulated as follows:

\begin{equation}
\label{eq:kd_optimization}
\begin{aligned}
 \{\boldsymbol{\Theta}_s^\star, \boldsymbol{\Phi}_s^\star\} 
     =&\underset{\substack{\boldsymbol{\Theta}, \boldsymbol{\Phi} }}{\arg \min } \ \frac{1}{P} \sum_{p=1}^P  \lambda_1 \| \mathcal{M}_{\boldsymbol{\Theta}}(\mathbf{A}_{\boldsymbol{\Phi}}^\top \mathbf{A}_{\boldsymbol{\Phi}} \mathbf{x}_p) -\mathbf{x}_p \|_2^2  \\ &+  \lambda_2 \mathcal{L}_{\text{DEC}} + \lambda_3 \mathcal{L}_{\text{ENC}}  + \tau \mathcal{R}_{\tau}(\mathbf{\Phi}_s) + \mu \mathcal{R}_{\mu}(\boldsymbol{\Theta}_s),\\
\end{aligned}
\end{equation}

The decoder KD loss function, $\mathcal{L}_{\text{DEC}}$, remains consistent across all three CI systems. This loss focuses on aligning the feature space of the student and teacher computational decoders. Since the teacher network contains clearer and more robust features—having been trained to reconstruct images with fewer degradations due to having {a} less constrained {encoder} than the student—its feature space can guide and improve the student's performance. To this end, we use the bottleneck layers, as they provide the most compact and informative representation of the input image. The decoder loss function is defined as:

\begin{equation}
\label{eq:FB_MRI_UNET_BOTTLENECK}
    \mathcal{L}_{\text{DEC}} = \frac{1}{B} \| 
    \mathbf{T}_b - \mathbf{S}_b
    \|_2^2,
\end{equation}

\noindent where $\mathbf{T}_b = \mathcal{M}_{\boldsymbol{\Theta}^\star_t}^{[b]}(\mathbf{A}_{\boldsymbol{\Phi}_t^\star}^\top \mathbf{A}_{\boldsymbol{\Phi}_t^\star} \mathbf{x})$ and $\mathbf{S}_b = \mathcal{M}_{\boldsymbol{\Theta}_s}^{[b]}(\mathbf{A}_{\boldsymbol{\Phi}_s}^\top \mathbf{A}_{\boldsymbol{\Phi}_s} \mathbf{x})$ represent the intermediate feature maps extracted from the bottleneck layers ($[b]$) of the teacher and student decoders, respectively. 

Since the encoder depends on the structure of each acquisition system (MRI, SPC, SD-CASSI), the encoder loss is defined for each CI system:

\subsubsection{Magnetic resonance imaging}

The encoder loss function proposed for MRI systems is:

\begin{equation} 
\mathcal{L}_{\text{ENC}} = \frac{1}{B} \left\| \mathbf{F}^{\text{H}} \boldsymbol{\Phi}_s^\top \boldsymbol{\Phi}_s \mathbf{F} \mathbf{x} - \mathbf{F}^{\text{H}}
{\boldsymbol{\Phi}_t^\star}^\top \boldsymbol{\Phi}_t^\star \mathbf{F} \mathbf{x} \right\|_2^2. 
\end{equation}

\noindent where $\mathbf{F}^{\text{H}}$ is the transpose conjugate of the Fourier transform matrix $\mathbf{F}$. This function encourages the student's $k-\text{space}$ undersampling mask $\boldsymbol{\Phi}_s$ to mimic the structure of the teacher's $k-\text{space}$ undersampling mask $\boldsymbol{\Phi}_t^\star$ by aligning the backprojected measurements from the student's and teacher's encoders in the image domain.

\subsubsection{Single-pixel camera}
For the SPC system, the encoder loss function is defined as:

\begin{equation}
    \mathcal{L}_{\text{ENC}} = \frac{1}{B} \left\| \mathbf{A}_{\mathbf{W}_s}^{\top} \mathbf{A}_{\mathbf{W}_s}-  \mathbf{A}_{\boldsymbol{\Phi}_t^\star}^{\top} \mathbf{A}_{\boldsymbol{\Phi}_t^\star} \right\|_2^2,
\end{equation}

\noindent where $\mathbf{A}_{\mathbf{W}s}$ is the student’s forward model with CAs before the binarization step, and $\mathbf{A}_{\boldsymbol{\Phi}_t^\star} \in \mathbb{R}^{m \times n}$ is the teacher's forward model, with the teacher employing real-valued CAs, $\boldsymbol{\Phi}_t^\star = \mathbf{W}_t^\star$. This loss encourages structural similarity in the student's forward model with the teacher's by aligning their Gram matrices, which capture the pairwise correlations of the CA patterns.

\subsubsection{Single disperser coded aperture snapshot spectral imager}

For the SD-CASSI system, the encoder loss function is expressed as:

\begin{equation}
    \mathcal{L}_{\text{ENC}}=\frac{1}{B}\| \mathbf{W}_s^\top \mathbf{W}_s - {\boldsymbol{\Phi}_t^\star}^\top {\boldsymbol{\Phi}_t^\star} \|_2^2,
\end{equation}

\noindent where $ \mathbf{W}_s $ is the student CA prior binarization, and $\boldsymbol{\Phi}_t^\star$ is the optimal teacher real-valued CA. This function promotes structural similarity between the Gram matrices of the student and teacher CAs.

\section{Simulations and Results}

Reconstruction performance was measured using the peak signal-to-noise ratio (PSNR) and structural similarity index (SSIM) \cite{SSIM_METRIC_PAPER}. The teacher and baseline models for all three CI systems were previously optimized according to equation \eqref{eq:e2e_optimization}. All students were trained under the guidance of their optimal teacher, corresponding to a less constrained CI system with a similar number of snapshots and a comparable nature of the {codification basis of the encoder}. For specific details on how the optimal teachers were selected, refer to the supplementary material (Section III).


Furthermore, the improvement in the encoder design was evaluated for each CI system. In MRI we utilized the learned undersampling mask with the baseline and student as fixed undersampling patterns to train a reconstruction network. We computed the condition number, singular value distribution, and gram matrices of the forward models for the SPC. In the SD-CASSI system, we computed the spectral band correlation of the optimized forward model and the magnitude of the Fourier transform of the designed CA.  Additional experiments, including computational requirements, loss function ablation studies, teacher system selection, and noise robustness evaluations, are detailed in the supplementary material. The repository containing the code is available at\footnote{\href{https://github.com/leonsuarez24/DKDCIS}{\textcolor{blue}{\textbf{https://github.com/leonsuarez24/DKDCIS}}}}. The implementation employs PyTorch \cite{PyTorch_BIB} for deep learning components and Colibri \footnote{\textcolor{blue}{\textbf{https://github.com/pycolibri/pycolibri}}} for optical system simulations.

The following subsections describe the simulations and results for each CI system employed.

\subsection{Magnetic resonance imaging}

\textbf{Training details:} The FastMRI single-coil knee dataset \cite{FastMRI_dataset}, obtained from  the DeepInverse library \cite{deepinv}, was used. It consists of magnetic resonance images of knees, each with a resolution of $320\times320$. The dataset includes $900$ training images, of which $810$ were used for training and $90$ for validation. The test set contains $73$ images. All images were resized to $256\times256$. The teacher, student, and baseline models were trained for $500$ epochs. The first 200 epochs constituted an unconstrained phase, where the acceleration factor regularization function in \eqref{eq:reg_tramittance} used a regularization parameter $\tau=1$, allowing the exploration of different undersampling schemes. The remaining 300 epochs formed a constraining and refining phase, with the regularization parameter set to $\tau=1\times 10^{15}$ to limit the mask $\boldsymbol{\Phi}$ to the given acceleration factor $AF$ and refine the decoder network training with the learned mask. The training was conducted with a $5\times10^{-4}$ learning rate and a batch size of $B=32$. The AdamW \cite{AdamW} optimizer was employed with a weight decay parameter of $1 \times 10^{-2}$. For the following results, we set $\lambda_1 = 0.1, \lambda_2 = 0.3$ for $AF_s=4$, $\lambda_1 = 0.3, \lambda_2 = 0.2$ for $AF_s=8$, and $\lambda_1 = 0.3, \lambda_2 = 0.5$ for $AF_s=16$. These values were chosen based on a hyperparameter search detailed in Section I of the Supplementary Material.

We compared the proposed KD approach with the traditional E2E optimization scheme for designing MRI systems. The teacher systems were previously optimized with an acceleration factor $AF_t = AF_s - 1$ {(for details of this selection, refer to Section III of the supplementary material)}. Five realizations of the student MRI systems and baselines were performed, with the average and standard deviation of these realizations reported. Table \ref{tab:MRI_RESULTS_TABLE} summarizes the obtained results. It can be observed that in all three evaluated acceleration factors, the student achieves better results than the baseline. Furthermore, the results indicate that all student models are more stable than the baseline, exhibiting lower standard deviations in the reconstruction metrics. 

\begin{figure*}[t!]
    \centering
    \includegraphics[width=0.9\textwidth]{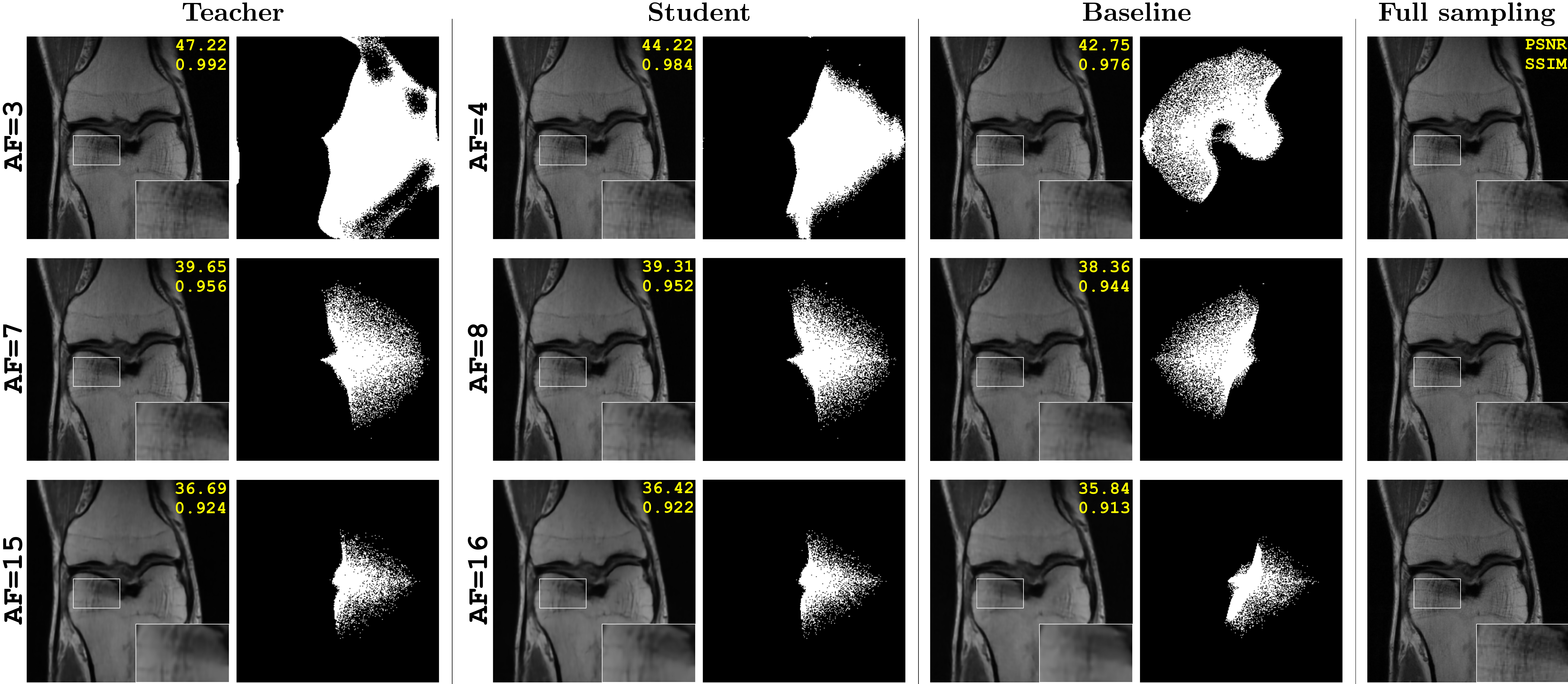}
    \caption{Reconstruction performance of the student and baseline MRI systems. The first column shows the teacher-optimized undersampling mask and its corresponding reconstruction. The second column presents the student-optimized mask and its reconstruction. The third column displays the baseline-optimized mask and its reconstruction, while the fourth column contains the ground truth image. The PSNR (dB) and SSIM metrics are reported in the upper-right corner of each reconstruction.}
    \label{fig:MRI_exps_U_net}
\end{figure*}

Figure \ref{fig:MRI_exps_U_net} shows visual results comparing the proposed method with the E2E baseline in terms of reconstruction performance and undersampling $k-\text{space}$ mask design. As the figure illustrates, the student's undersampling mask effectively imitates the teacher's undersampling mask across different acceleration factors, due to the use of the $\mathcal{L}_{\text{ENC}}$ function, yielding better reconstructions than the baseline, getting improvement by up to $1.47$ (dB) in PSNR. Additionally, zoom-in sections demonstrate the student model's superior ability to recover high-frequency details and preserve fine structural information than the baseline.

\begin{table*}[t!]
\centering
\caption{Reconstruction performance of student and baseline MRI systems for $AF \in \{4, 8, 16\}$. The teacher system has an acceleration factor $AF_t$ that is one unit lower than the corresponding student $(AF_t = AF_s - 1)$. Best results are shown in bold}
\label{tab:MRI_RESULTS_TABLE}
\begin{tabular}{ccccccc}
\hline

\multirow{2}{*}{$AF_s$} & \multicolumn{2}{c}{Teacher $(AF_t = AF_s - 1)$} & \multicolumn{2}{c}{Student}                 & \multicolumn{2}{c}{Baseline}                \\ \cline{2-7} 
                        & PSNR $\uparrow$        & SSIM $\uparrow$        & PSNR $\uparrow$      & SSIM $\uparrow$      & PSNR $\uparrow$      & SSIM $\uparrow$      \\ \hline
4                       & $45.48$                & $0.985$                & $\mathbf{42.82 \pm 0.12}$     & $\mathbf{0.974 \pm 0.0004}$   & $41.52 \pm 0.28$     & $0.962 \pm 0.0025$   \\
8                       & $39.07$                & $0.938$                & $\mathbf{38.62 \pm 0.06}$     & $\mathbf{0.932 \pm 0.0006}$   & $38.26 \pm 0.22$     & $0.928 \pm 0.0024$   \\
16                      & $36.57$                & $0.905$                & $\mathbf{36.26 \pm 0.17}$     & $\mathbf{0.901 \pm 0.0022}$   & $35.96 \pm 0.22$     & $0.896 \pm 0.0027$   \\ \hline

\end{tabular}
\end{table*}

\begin{figure}[!t]
    \centering
    \includegraphics[width=0.85\linewidth]{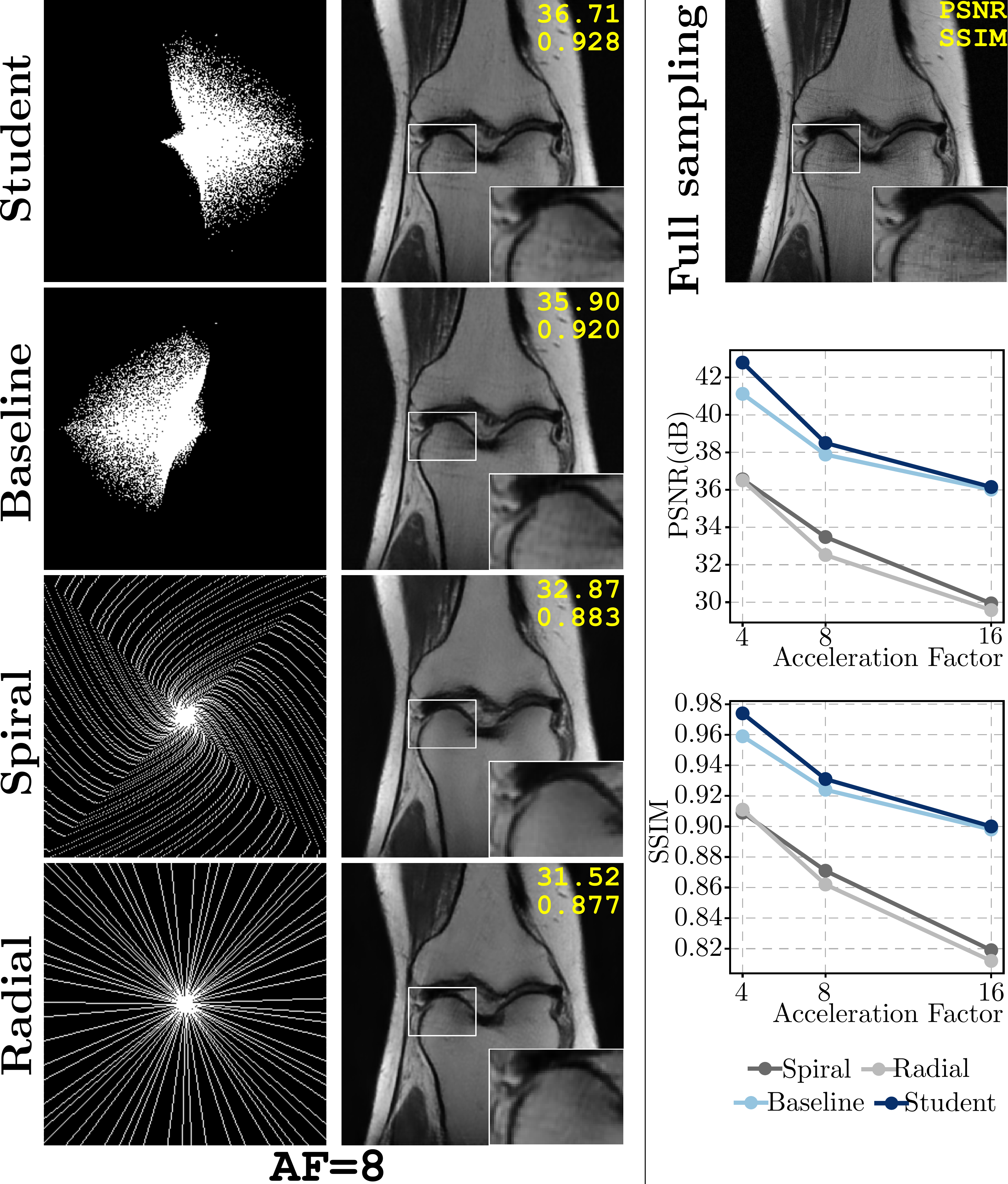}
    \caption{Comparison of the student MRI system with the baseline and common $k$-space undersampling masks (spiral and radial). On the left, a visual comparison for $AF=8$ is presented, with PSNR (dB) and SSIM metrics displayed in the upper-right corner of each reconstruction. On the right, a plot comparison for $AF \in \{4, 8, 16\}$ is shown.}
    \label{fig:SPIRAL_RADIAL}
\end{figure}

To evaluate the improvements in the design of the {encoder}, the optimized $k-\text{space}$ undersampling masks for the baseline and student systems were extracted, fixed its weights, and used to reconstruct magnetic resonance images from the undersampled measurements using a reconstruction neural network with the same architecture as the employed U-Net. Additionally, the reconstruction performance of the recovery network using the student’s learned $k-\text{space}$ undersampling masks was compared to traditional undersampling mask patterns, such as spiral and radial, from \cite{CIRCUS_MRI}. Figure \ref{fig:SPIRAL_RADIAL} presents visual results of the reconstruction performance for the trained neural network with the fixed learned masks of the student and baseline, as well as with the spiral and radial patterns, all with $AF=8$. These results show that the network trained with the student’s learned mask outperforms the baseline as well as the spiral and radial patterns. Zoomed-in results reveal that the network trained with the student’s mask captures finer high-frequency features, leading to improved reconstruction. Additionally, a plot for $AF \in \{4, 8, 16\}$ demonstrates that the network trained with the student mask consistently outperforms the networks trained with the baseline, spiral, and radial masks across all acceleration factors.

\subsection{Single-pixel camera}

\textbf{Training details:} The FashionMNIST dataset \cite{fashionmnist} was employed. It consists of $60,000$ training and $10,000$ testing images of size $28\times 28$ containing $10$ classes of clothing. The training dataset was divided into $50,000$ images for training and $10,000$ for validation. All images were resized to $32\times32$. The training was performed for $50$ epochs using the Adam optimizer \cite{Adam} with a learning rate of  $5 \times 10^{-4}$ and a batch size of $B=64$ images.

We compared the proposed KD approach with the traditional E2E optimization scheme for designing SPC systems. All teacher systems were previously trained with $\gamma_t = \gamma_s$ with real-valued CAs. Five realizations of the student and baseline models were evaluated, with average and standard deviation reported. An additional comparison was made using the traditional Hadamard basis for the CA patterns \cite{zhang2017hadamard}, with the computational decoder trained accordingly. Table \ref{tab:reconstruction_SPC} shows the results, where the student system outperforms both the baseline (by up to $1.05$ (dB)) and the Hadamard basis across all compression ratio configurations. Additionally, the student models, in almost all cases, demonstrate greater stability than the baseline, showing lower standard deviations in the reconstruction metrics. 

\begin{figure*}[!t]
    \centering
    \includegraphics[width=0.9\linewidth]{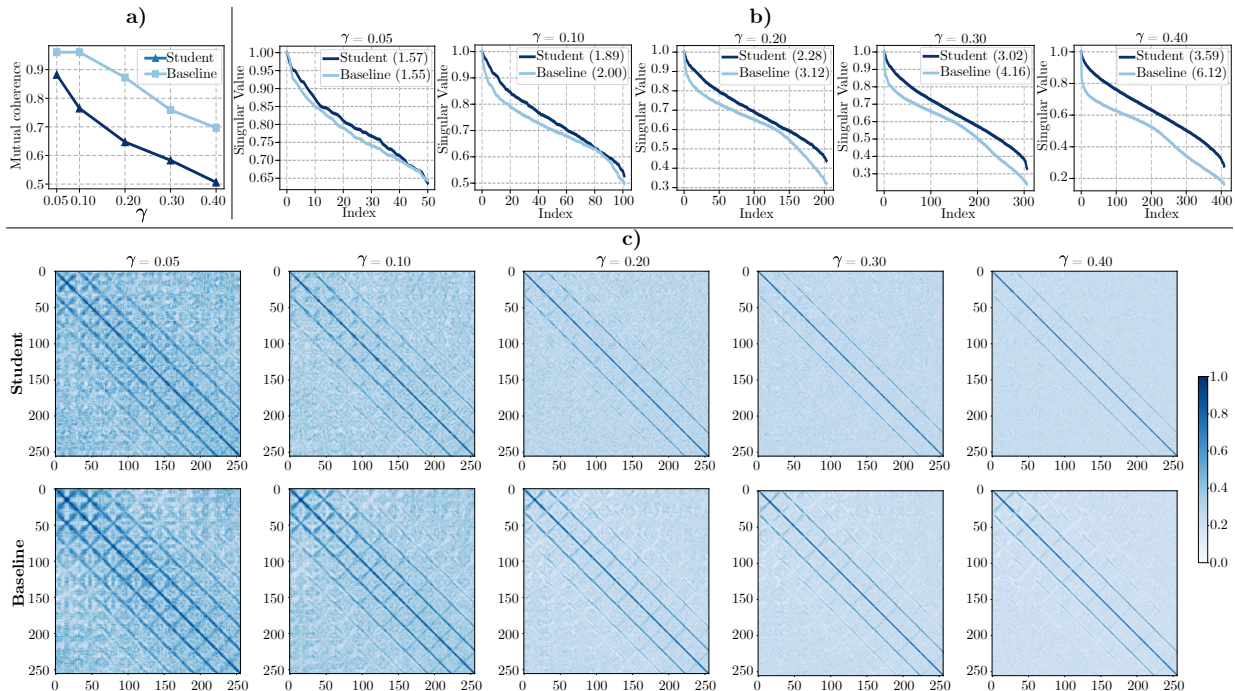}
    \caption{(a) Mutual coherence of the student and baseline SPC sensing matrices for various compression ratios. (b) Distribution of normalized singular values for the student and baseline SPC matrices at different compression ratios, with their condition numbers indicated next to the respective labels. (c) Visualization of a $256 \times 256$ section of the $\mathbf{A}_{\boldsymbol{\Phi}}^\top \mathbf{A}_{\boldsymbol{\Phi}}$ matrices for the student's SPC and the baseline's sensing matrices.}
    \label{fig:GRAMMS_MATRICES}
\end{figure*}

To evaluate the improvement in the codification of the CAs learned by the student, we present the mutual coherence $\mu(\mathbf{A}) = \max_{i \neq j} \left| \langle \mathbf{a}_i, \mathbf{a}_j \rangle \right|$ of the student and baseline forward model matrices across all evaluated compression ratios in Figure \ref{fig:GRAMMS_MATRICES}a. These results show that the columns of $\mathbf{A}_{\boldsymbol{\Phi}_s^\star}$ (the student's sensing matrix) are less correlated with each other compared to the columns of $\mathbf{A}_{\boldsymbol{\Phi}^\star}$ (the baseline's sensing matrix). This indicates that the student can collect a more diverse set of measurements than the baseline. Additionally,  Figure \ref{fig:GRAMMS_MATRICES}b shows the distribution of the normalized singular values of the student and baseline forward model operators. The student's matrix exhibits larger minimal singular values, leading to a lower condition number $\kappa(\mathbf{A}) = \frac{\sigma_{\text{max}}(\mathbf{A})}{\sigma_{\text{min}}(\mathbf{A})}$ compared to the baseline,  showing that the student’s system is more stable and robust than the baseline. Furthermore,  Figure \ref{fig:GRAMMS_MATRICES}c provides visual results comparing sections of the Gram matrices $\mathbf{A}_{\boldsymbol{\Phi}_s^\star}^\top \mathbf{A}_{\boldsymbol{\Phi}_s^\star}$ for the student and $\mathbf{A}_{\boldsymbol{\Phi}^\star}^\top \mathbf{A}_{\boldsymbol{\Phi}^\star}$ for the baseline. The student's Gram matrix demonstrates stronger linear independence due to its off-diagonal elements having lower values compared to the baseline’s Gram matrix. This suggests that the student’s matrix provides more distinct and non-redundant information, contributing to improved performance in the reconstruction due to measurement diversity and stability of the sensing matrix.

\begin{table*}[!t]
\centering
\caption{Reconstruction performance of student and baseline SPC systems for $\gamma \in \{0.05, 0.1, 0.2, 0.3, 0.4\}$. The teacher system has the same compression ratios as the student $\gamma_t=\gamma_s$, however, it uses real-valued coded apertures. Best results are shown in bold}
\label{tab:reconstruction_SPC}
\begin{tabular}{cllcccccc}
\hline
                           & \multicolumn{2}{c}{Teacher}                                 & \multicolumn{2}{c}{Student}          & \multicolumn{2}{c}{Baseline}  & \multicolumn{2}{c}{Hadamard}          \\ \cline{2-9} 
\multirow{-2}{*}{$\gamma$} & \multicolumn{1}{c}{PSNR $\uparrow$}     & \multicolumn{1}{c}{SSIM $\uparrow$}     & PSNR  $\uparrow$           & SSIM  $\uparrow$            & PSNR    $\uparrow$         & SSIM      $\uparrow$  & PSNR    $\uparrow$         & SSIM      $\uparrow$        \\ \hline
0.05                       & 25.59 & 0.915 & $\mathbf{24.50 \pm 0.04}$ & $\mathbf{0.894 \pm 0.001}$ & $24.03 \pm 0.03$  & $0.887 \pm 0.002$ & $18.80 \pm 0.03$  & $0.745 \pm 0.004 $\\
0.1                        & 27.84 & 0.944 & $\mathbf{26.07 \pm 0.04}$ & $\mathbf{0.920 \pm 0.001}$  & $25.53 \pm 0.09$ & $0.907 \pm 0.009$ & $19.78 \pm 0.03$  & $0.793 \pm 0.006$\\
0.2                        & 30.90 & 0.972 & $\mathbf{27.93 \pm 0.06}$ & $\mathbf{0.945 \pm 0.001}$ & $27.06 \pm 0.08$  & $0.936 \pm 0.002$ & $22.60 \pm 0.03$  & $0.867 \pm 0.013$\\
0.3                        & 33.29 & 0.978 & $\mathbf{29.11 \pm 0.08}$ & $\mathbf{0.955 \pm 0.002}$ & $28.06 \pm 0.21$  & $0.947 \pm 0.003$ & $ 24.95 \pm 0.05$  & $0.911\pm 0.006 $\\
0.4 			   & 34.57 & 0.987 & $\mathbf{29.54 \pm 0.06}$ & $\mathbf{0.961 \pm 0.001}$ & $28.81 \pm 0.2$ & $0.955 \pm 0.002$  & $26.83 \pm  0.08$  & $0.938\pm 0.004$\\ \hline
\end{tabular}
\end{table*}

\subsection{Single disperser coded aperture snapshot spectral imager}

\textbf{Training details:} The ARAD-1K dataset \cite{ARA1K} was used for multi-spectral image reconstruction. It contains $950$ images, each of size $31 \times 512 \times 512$ in the 400-700 nm range. The images were resized to $256\times 256$ and eight equidistant spectral bands were selected from the original 31 bands. Four non-overlapping patches were extracted from each image, resulting in $3,600$ patches for training, $125$ for validation, and $75$ for testing, with each patch sized $8 \times 128 \times 128$. The AdamW \cite{AdamW} optimizer with a weight decay of $1\times 10^{-2}$ was employed. The teacher, student, and baseline systems were trained for $500$ epochs with a learning rate of $5 \times 10^{-4}$, using a batch size of $B=32$.

The proposed approach was compared to traditional E2E optimization for a SD-CASSI system with a binary-valued CA and one-snapshot configuration. The teacher system was pretrained using a real-valued CA with the same one-snapshot configuration. Five realizations of the student and baseline models were conducted, with the average and standard deviation reported. Additionally, we compared the reconstruction performance of the network using a SD-CASSI with a Blue Noise CA \cite{Blue_NOISE} with one snapshot. Table \ref{tab:CASSI_EXPS} summarizes the results, demonstrating that the student model  outperforms the baseline (by up to $1.53$ dB in PSNR) and Blue Noise CA (by up to $1.56$ (dB) in PSNR).  

\begin{figure}[!t]
    \centering
    \includegraphics[width=\linewidth]{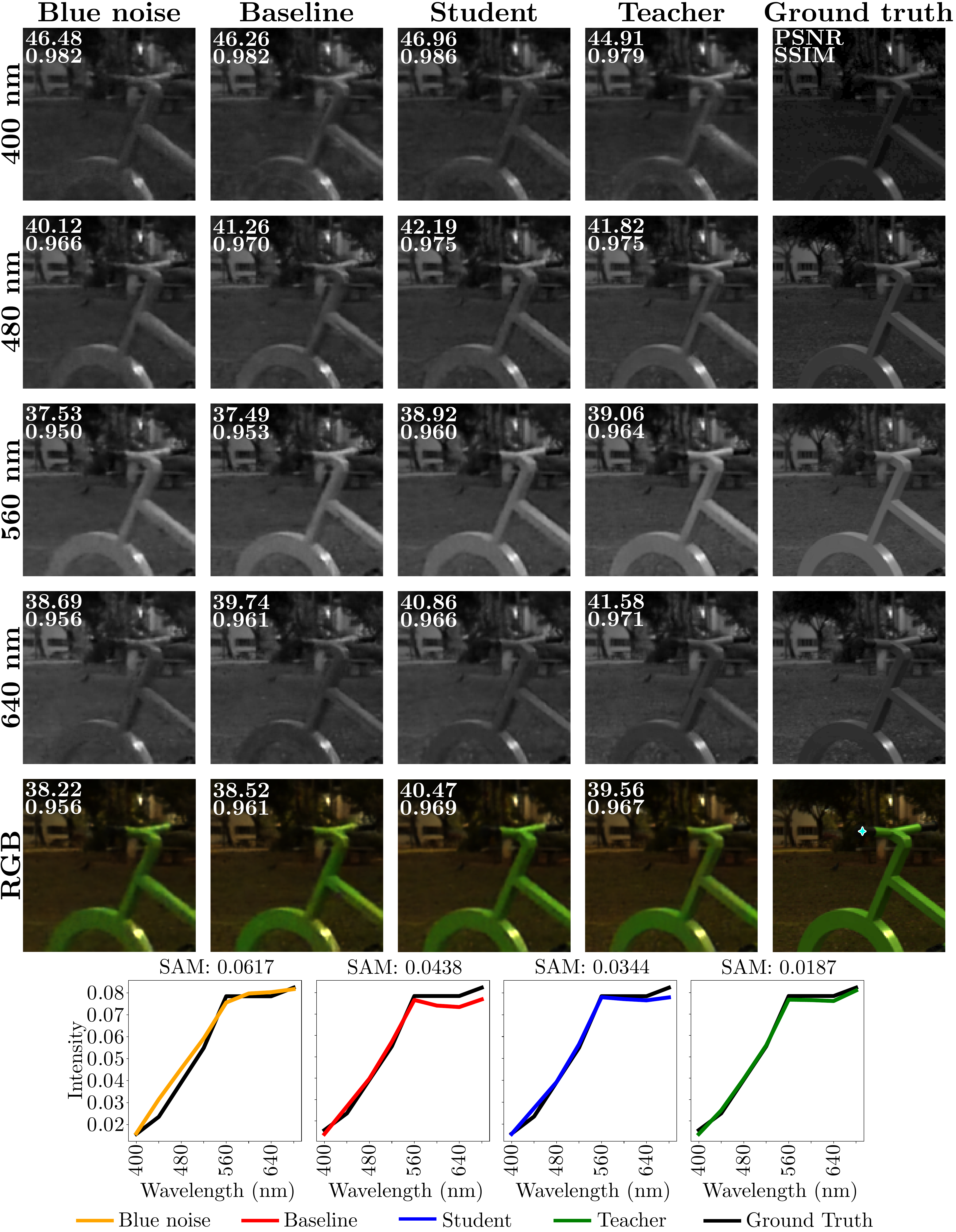}
    \caption{Reconstruction performance of the SD-CASSI system, reported in PSNR (dB) and SSIM, for the teacher, student, baseline, and Blue Noise models. Results are shown for all eight spectral bands, detailing the PSNR and SSIM for each band. The overall performance across all eight spectral bands is: teacher (PSNR: 40.30 (dB), SSIM: 0.971), student (PSNR: 40.90 (dB), SSIM: 0.971), baseline (PSNR: 39.43 (dB), SSIM: 0.965), and Blue Noise (PSNR: 38.86 (dB), SSIM: 0.960). Additionally, the spectral profiles for a point in the RGB ground truth image are presented, along with the SAM metric. }
    \label{fig:VISUAL_CASSI}
\end{figure}

Figure \ref{fig:VISUAL_CASSI} presents visual reconstructions of four spectral bands of the eight, comparing the proposed KD approach with E2E optimization and the use of a Blue Noise CA as a fixed encoder. The results are shown in terms of PSNR and SSIM. The results demonstrate that the student model consistently outperforms both the baseline and Blue Noise CA across all bands, and in nearly all bands, it even surpasses the teacher model. For the multi-spectral image, the student achieves a PSNR of 40.90 (dB), surpassing the teacher's 40.30 (dB). The student also outperforms the baseline and Blue Noise CA, with improvements of 1.47 (dB) over the baseline and 2.04 (dB) over the Blue Noise CA. Also, spectral profiles were evaluated at the RGB ground truth image point. The Spectral Angle Mapper (SAM) metric \cite{SAM_METRIC} was used to quantify spectral similarity. The results show that the student system achieves a SAM score of 0.0344, outperforming the baseline score of 0.0433 and the Blue Noise CA score of 0.0617 (lower values indicate better similarity). 


\begin{table}[!t]
\centering 
\caption{Reconstruction performance of the student, baseline, Blue Noise CA, and teacher SD-CASSI systems. The baseline, student, and Blue Noise models use a binary-valued CA, while the teacher uses a real-valued CA.}
\label{tab:CASSI_EXPS}
\begin{tabular}{ccc}
\hline
         & PSNR  $\uparrow$ & SSIM $\uparrow$ \\ \hline
Teacher  & $37.44$ & $0.965$   \\
Student  & $\mathbf{36.89 \pm 0.01}$& $\mathbf{0.958 \pm 0.001}$  \\
Baseline & $35.36 \pm 0.21$   & $0.943 \pm 0.002$  \\ 
Blue Noise & $35.33\pm0.03$   & $0.942\pm0.001$  \\ 
\hline
\end{tabular}
\end{table}

\begin{figure}[!t]
    \centering
    \includegraphics[width=0.91\linewidth]{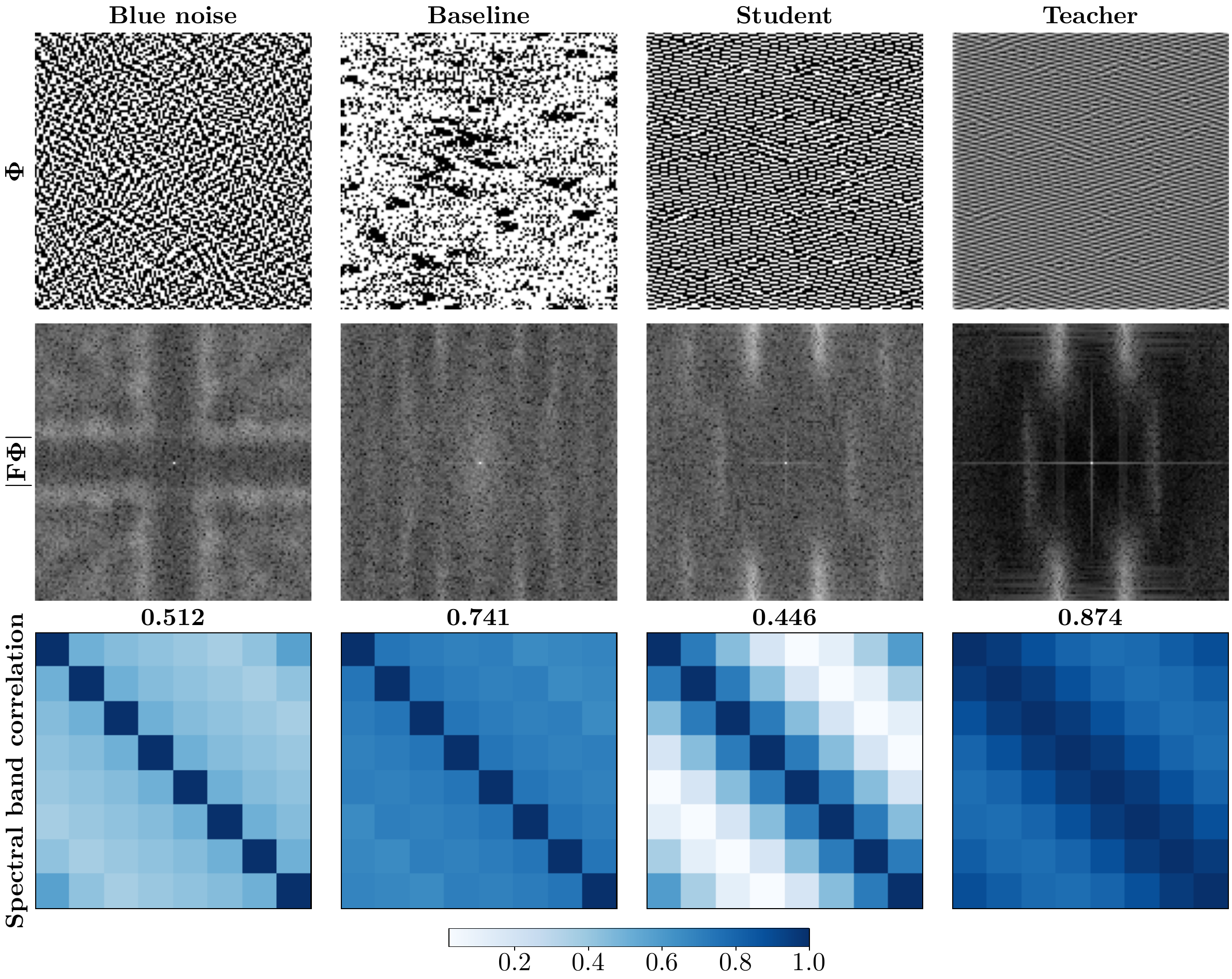}
    \caption{First row shows the learned CA with the teacher, student and baseline models, and the used Blue Noise CA. Second row shows the magnitude of the Fourier Transform of the CAs of the teacher, student, baseline and Blue Noise. Third row shows the spectral band correlation matrix of the student}
    \label{fig:CAS_CASSI_FFT}
\end{figure}

To evaluate the improvements in the design of the {encoder} for the SD-CASSI system using the proposed approach, we computed the spectral band correlation matrix $\mathbf{G} \in \mathbb{R}^{L \times L}$. This matrix quantifies the correlation between the spectral bands, where low correlation values indicate incoherent spectral sampling. The matrix is obtained by decomposing the SD-CASSI sensing matrix $\mathbf{A}_{\boldsymbol{\Phi}} \in \mathbb{R}^{N(M+L-1)\times NML}$ into $L$ blocks, such that $\mathbf{A}_{\boldsymbol{\Phi}} = [\mathbf{A}_1, \mathbf{A}_2, \dots, \mathbf{A}_L]$, where each $\mathbf{A}_i \in \mathbb{R}^{N(M+L-1)\times NM}$ corresponds to the sensing matrix for the $i$-th spectral band, containing the diagonalized CA shifted by $i \cdot N$ positions. The $(i,j)$-th entry of the spectral band correlation matrix $\mathbf{G}$ is computed as the inner product of the submatrices $\mathbf{A}_i$ and $\mathbf{A}_j$, given by $\mathbf{G}_{i,j} = \operatorname{Tr}(\mathbf{A}_i^\top \mathbf{A}_j)$. Additionally, we computed the Fourier transform magnitude of the CAs $\mathbf{F}\boldsymbol{\Phi}$. Figure \ref{fig:CAS_CASSI_FFT} illustrates both the spectral band correlation matrix and the Fourier transform magnitude metrics for the baseline, teacher, student, and Blue Noise CAs. The results show that the teacher system focuses on very specific low and high frequencies, discarding the rest of the spectrum. The student system displays a similar frequency response to the teacher but with a broader emphasis across the spectrum, particularly enhancing both low and high frequencies, while also prioritizing certain regions. The baseline CA shows a more uniform behavior, with higher-frequency details present but not as dominantly represented as in the teacher or student CAs. The Blue Noise CA provides a more uniform encoding of both low and high frequencies. However, it’s not as high-frequency focused as a teacher, nor as low-frequency concentrated as student or baseline. Furthermore, the spectral band correlation matrix results show that the student CA exhibits lower coherence than the baseline and Blue Noise CAs, indicating that it is less spectrally correlated. This lower correlation suggests that the student CA acquires more diverse and less spectrally redundant measurements compared to the other two. Additionally, the student CA’s average correlation of 0.446 is lower than that of the Baseline (0.741) and Blue Noise (0.512). The higher average spectral band correlation seen in the teacher system can be attributed to the relaxation of the binary constraint. In other words, allowing the teacher CA to use real-valued intensity levels, rather than strictly binary values, leads to measurements that are more closely related spectrally.

\section{Discussion \& future work}
The proposed approach is designed as an alternative to traditional E2E optimization for the design of any CI system {by reinterpreting the concept of KD within CI systems. Instead of using a large and complex neural network as a teacher, it uses a less constrained, high-performance CI system, and the student corresponds to a more constrained CI system}. In this work, we validated it on three representative CI systems: MRI, the SPC, and the SD-CASSI. {Future work can address limitations of the proposed approach, such as the need for hyperparameter tuning of the KD loss functions, as these are not the same across all CI modalities,  and determining the criteria for selecting an optimal teacher to guide the student. Further research could also} explore the design of other CI systems, such as seismic imaging, phase retrieval, and diffractive optical imaging systems. Additionally, while only a few relaxations of the teacher’s {encoder} were explored, future research could investigate less conventional parameterization found useful in neural networks and inverse problems such as complex  \cite{lee2022complex} or hypercomplex numbers \cite{bojesomo2024deep,jacome2024invitation}, or employ a different CI system as the teacher, distinct from the student. Furthermore, this work addresses the image reconstruction task; future work could explore other tasks, such as segmentation, classification, and depth estimation. Additionally, this approach can be useful for designing recovery-only schemes without designing the CE, which can be useful for a wide range of imaging inverse problems such as super-resolution, computed tomography, (non-) blind deconvolution, and denoising \cite{ongie2020deep}.

\section{Conclusion}

We propose a general-purpose, novel approach for the design of CI systems based on KD, which is competitive with traditional E2E optimization. We validated this approach on three representative CI systems—MRI, SPC, and SD-CASSI—through multiple experiments. The results demonstrate that by relaxing the original optimization problem and solving it, the resulting teacher model provides valuable knowledge that is transferred via two types of loss functions: encoder and decoder loss functions, which enhance the recovery performance of the student model. This approach leads to improved reconstruction performance and better encoder design than traditional E2E optimization, achieving gains of up to 1.47 (dB) in MRI, 1.05 (dB) in SPC, and 1.53 (dB) in the SD-CASSI. Additionally, improvements in encoder design were observed, including better condition numbers, mutual coherence, and linear independence in the SPC, better spectral band correlation in SD-CASSI systems, as well as enhanced reconstruction performance in MRI when using the learned undersampling mask as a fixed encoder within reconstruction networks. Furthermore, validation experiments confirm the robustness of the proposed method under noisy measurement conditions.




\author{Leon Suarez-Rodriguez~\IEEEmembership{Student Member,~IEEE,}, Roman Jacome~\IEEEmembership{Student Member,~IEEE,}, Henry Arguello,~\IEEEmembership{Senior Member,~IEEE,}}



\bibliographystyle{IEEEtran}
\bibliography{bibliography}







\maketitlesupplementary
\setcounter{section}{0}
\section{Hyperparameter tuning}

This section presents the results of the hyperparameter tuning for the regularization terms of the KD loss function ($\lambda_1, \lambda_2$). A grid search was conducted for each CI system, evaluating $(\lambda_1, \lambda_2) \in \{0.1, 0.2, 0.3, 0.4, 0.5, 0.6, 0.7, 0.8\} \times \{0.1, 0.2, 0.3, 0.4, 0.5, 0.6, 0.7, 0.8\}$ with the constraint that $\lambda_1 + \lambda_2 < 1$ to prevent negative values for $\lambda_3$. 

\begin{figure*}[!t]
    \centering
    \includegraphics[width=\textwidth]{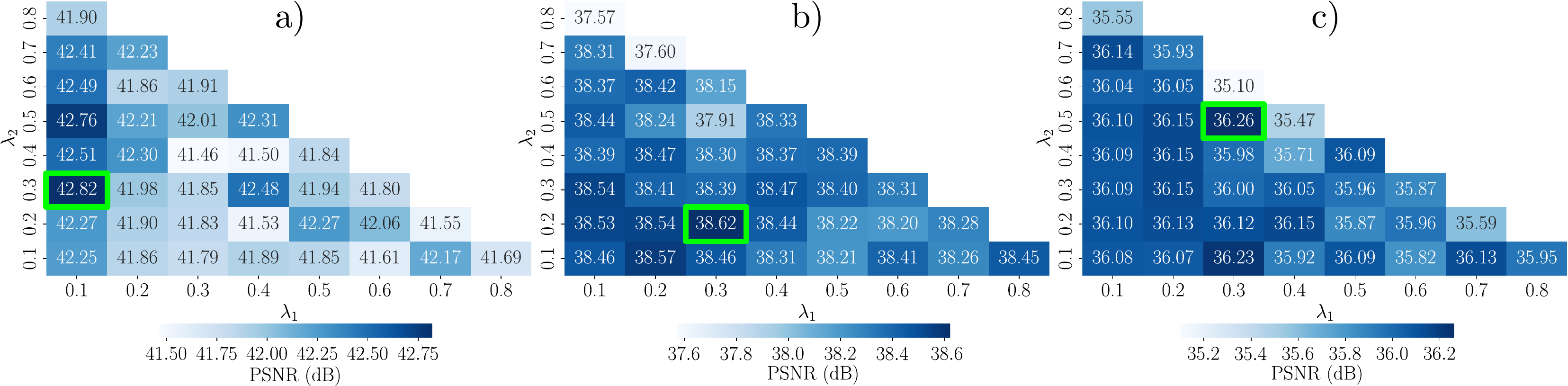}
    \caption{Grid search for hyperparameter tuning of $\lambda_1$ and $\lambda_2$ for MRI students with acceleration factors a) $AF_s=4$, b) $AF_s=8$, and c) $AF_s=16$, using a teacher model with $AF_t = AF_s - 1$ for each case. Results are reported in PSNR.}
    \label{fig:hyperparameter_tunning}
\end{figure*}

Figure \ref{fig:hyperparameter_tunning} shows the results of this search for MRI, with the tuning performed for each acceleration factor of the student, i.e., $AF_s \in \{4, 8, 16\}$, where the teacher model used $AF_t = AF_s - 1$ for each corresponding $AF_s$. We observed the following trends: for $AF_s = 4$, the optimal set was $\{\lambda_1=0.1, \lambda_2=0.3\}$ with $\lambda_3 =0.6$, indicating that the encoder loss function $\mathcal{L}_{\text{ENC}}$ plays a dominant role, with the student's performance heavily relying on mimicking the teacher's undersampling mask. For $AF_s = 8$, the optimal set was $\{\lambda_1=0.3, \lambda_2=0.2\}$ with $\lambda_3=0.5$, while for $AF_s = 16$, the optimal set was $\{\lambda_1=0.3, \lambda_2=0.5\}$ with $\lambda_3=0.2$. As the acceleration factor increases, the number of available points for subsampling the $k-\text{space}$ decreases, shifting the focus of the loss function more towards knowledge transfer in the decoder and less on the encoder.

For SPC, the optimal values were found to be $\lambda_1 = \lambda_2 = 0.1$, leading to $\lambda_3 = 0.8$. This configuration indicated that the student's performance was primarily influenced by minimizing the encoder loss by aligning the student's gram matrix of the forward operator before binarization $\mathbf{A}_{\mathbf{W}_s}^{\top} \mathbf{A}_{\mathbf{W}_s} $ with the teacher's gram matrix $\mathbf{A}_{\boldsymbol{\Phi}_t^\star}^{\top} \mathbf{A}_{\boldsymbol{\Phi}_t^\star}$.

For the SD-CASSI system, a similar behavior was observed as with the SPC. The encoder loss function played a more significant role in the performance than the other losses. However, the decoder loss did not contribute to improving performance as much as the encoder loss. The optimal configuration was found to be $\lambda_1 = 0.1$, $\lambda_2 = 0$, and $\lambda_3 = 0.9$, with the system's improvement heavily relying on aligning the student's gram matrix of the CAs before binarization $\mathbf{W}_s^\top \mathbf{W}_s$ with the teacher's real-valued CAs gram matrix $ {\boldsymbol{\Phi}_t^\star}^\top {\boldsymbol{\Phi}_t^\star}$.

\section{Noise robustness}

\begin{figure}[!t]
    \centering
    \includegraphics[width=0.8\linewidth]{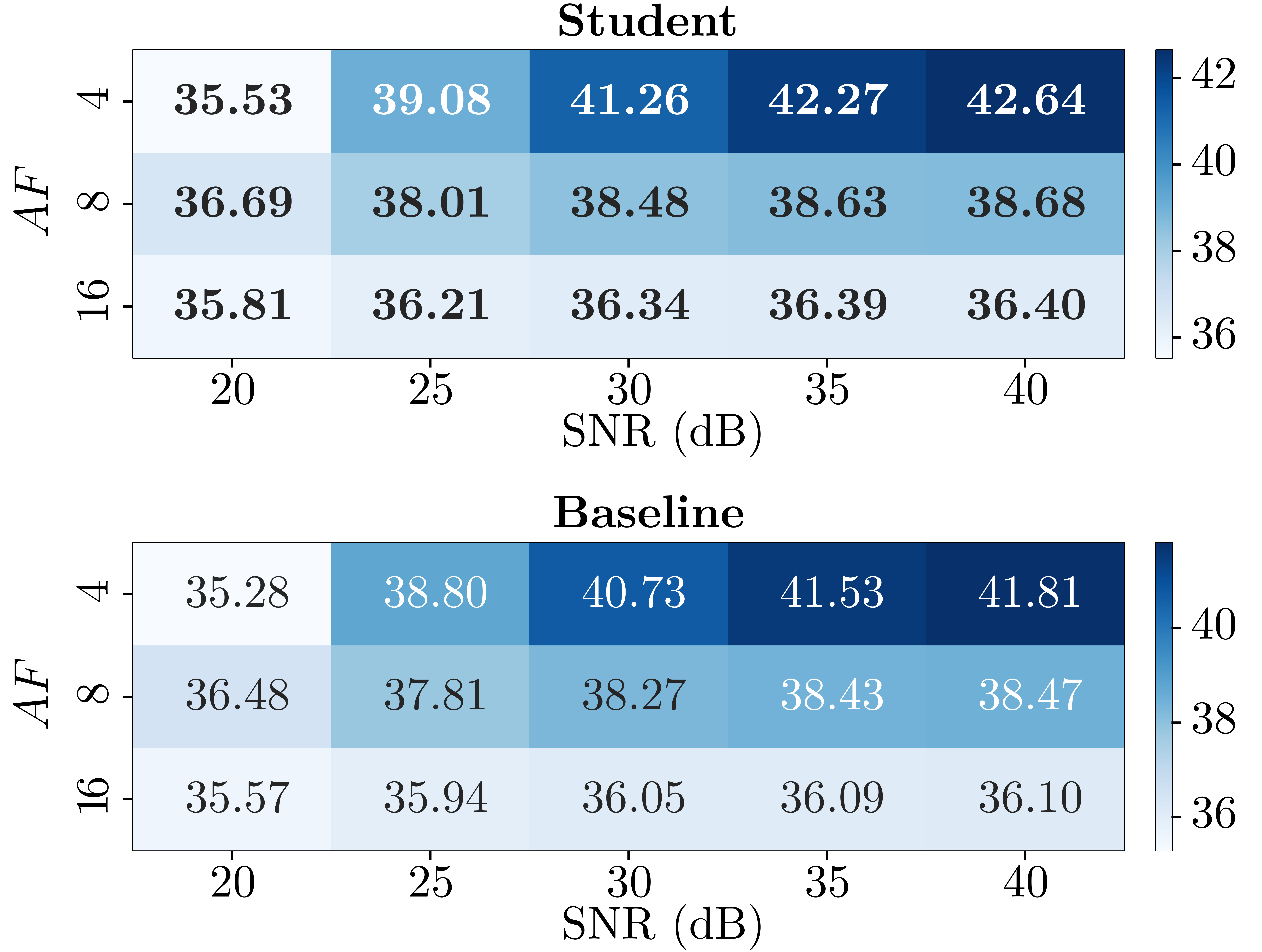}
    \caption{Reconstruction performance of the student and baseline models under varying noise levels for MRI systems in terms of PSNR.  Additive Gaussian noise with SNR values ranging from 20 dB to 40 dB was added to the measurements.}
    \label{fig:MRI_noise}
\end{figure}

Here, we evaluate the performance of the proposed approach in the presence of noise in the measurements. The analysis considers signal-to-noise ratios (SNRs) ranging from 20 dB to 40 dB, with increments of 5 dB. Figure \ref{fig:MRI_noise} presents the reconstruction results in terms of PSNR for MRI, showing that the student consistently outperforms the baseline across all noise configurations. For an acceleration factor of $AF=4$, both the student and baseline models exhibit the lowest PSNR values. This is primarily because the computational decoder at this acceleration factor was trained with less severe degradations compared to the decoders at higher acceleration factors. Additionally, although the student shows improved performance under noisy conditions, the overall gains are limited. This can be attributed to the fact that the computational decoders were trained on noise-free data. The performance could potentially be enhanced by incorporating noise into the $k-\text{space}$ measurements during the training phase. 

\begin{figure}
    \centering
    \includegraphics[width=0.8\linewidth]{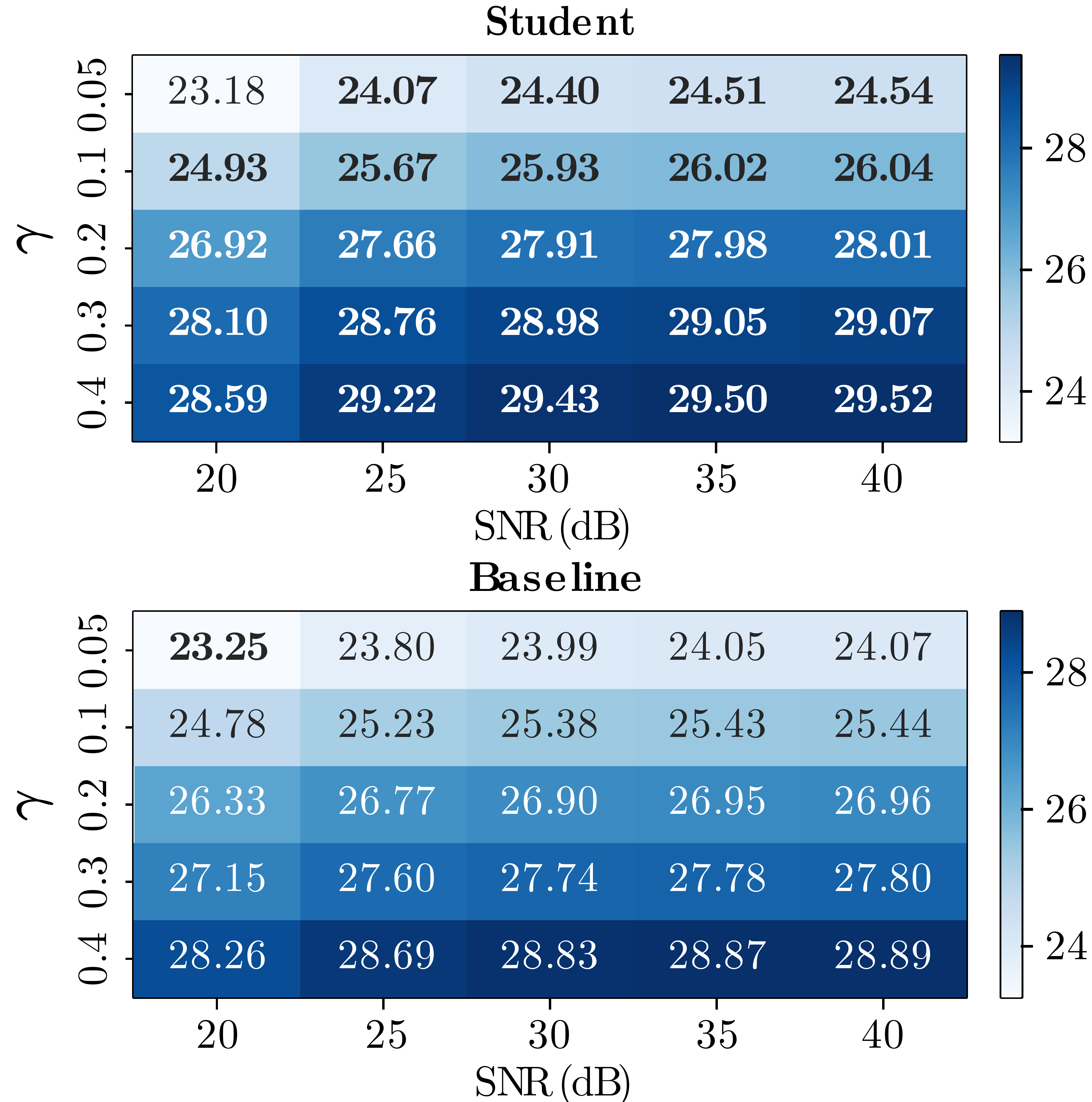}
    \caption{Reconstruction performance of the student and baseline models under varying noise levels for SPC systems in terms of PSNR.  Additive Gaussian noise with SNR values ranging from 20 dB to 40 dB was added to the measurements.}
    \label{fig:SPC_noise}
\end{figure}

Figure \ref{fig:SPC_noise} illustrates the reconstruction performance of the student and baseline models for SPC under noise conditions. In all cases, the student outperforms the baseline, except for $\gamma=0.05$ with a noise level of 20 dB. This exception may be attributed to the student having a slightly higher condition number $(1.57)$ compared to the baseline $(1.55)$, as shown in Figure 6b in the main text.

\section{Selection of the best teacher}

This section aims to determine which teacher CI system provides the most effective knowledge to the student. To achieve this, various teacher configurations were evaluated in the three CI systems.

\begin{figure}[!t]
    \centering
    \includegraphics[width=0.8\columnwidth]{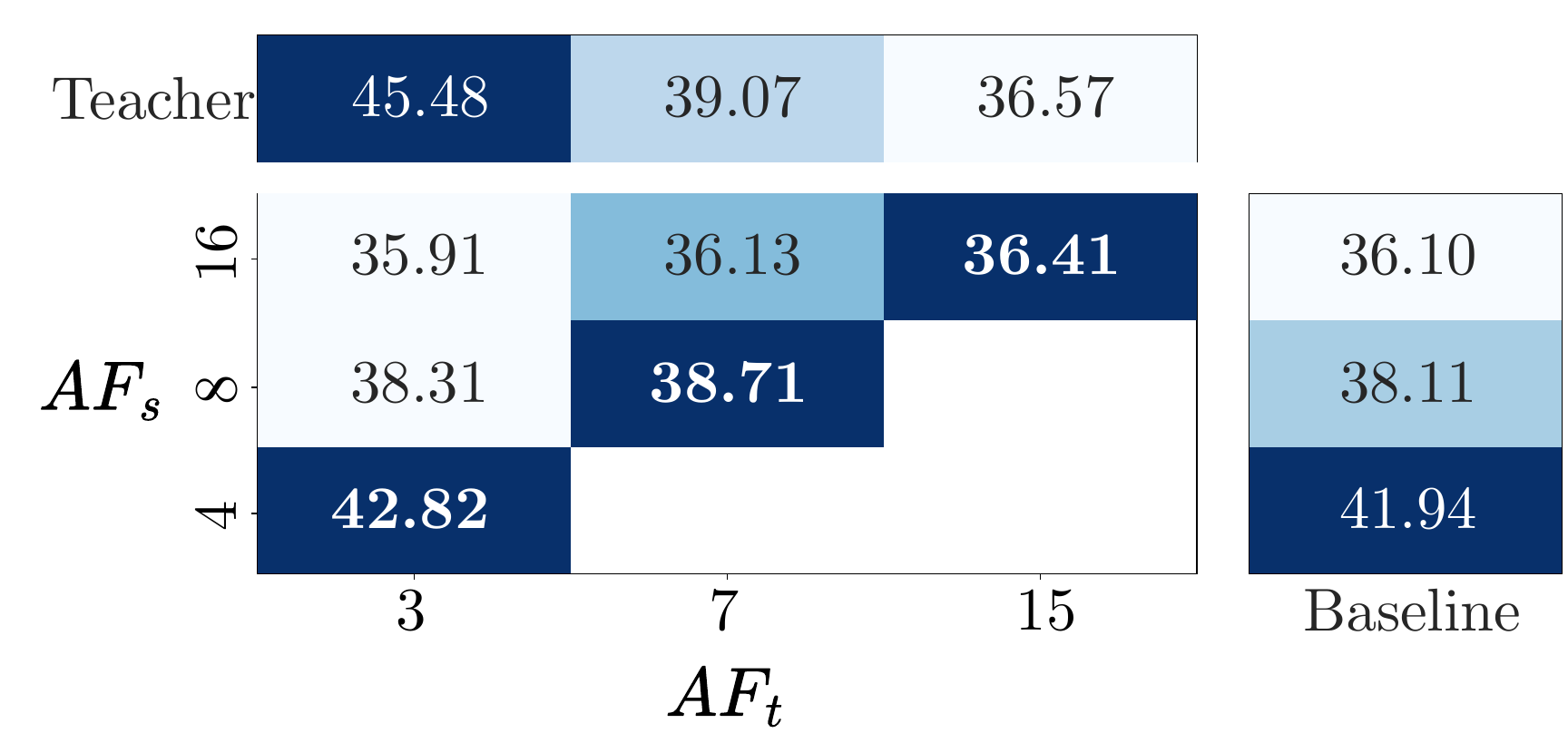}
    \caption{Performance of the student models when distilled from teacher systems with different acceleration factors ($AF_t \in \{3, 7, 15\}$) for $AF_s \in \{4, 8, 16\}$, with $AF_t < AF_s$. Results are reported in PSNR.}
    \label{fig:ABLATION_TEACHERS_MRI}
\end{figure}

In MRI, teachers with $AF_t \in \{3, 7, 15\}$ were used to distill students with $AF_s \in \{4, 8, 16\}$, under the condition that $AF_t < AF_s$. Figure \ref{fig:ABLATION_TEACHERS_MRI} presents the results, which show that the closer the teacher’s acceleration factor is to the student’s (while satisfying $AF_t < AF_s$), the better the student’s performance. This is partly due to the encoder loss function, which encourages the student’s $k-\text{space}$ undersampling mask to imitate the teacher’s mask. When the teacher and student have similar acceleration factors, the structure of the teacher’s mask is easier for the student to replicate, as the available sampling points in $k-\text{space}$ are more comparable. However, when the acceleration factors differ significantly, the difference in available sampling points increases, making it more challenging for the student to effectively imitate the teacher’s mask.

\begin{figure}[!t]
    \centering
    \includegraphics[width=0.85\linewidth]{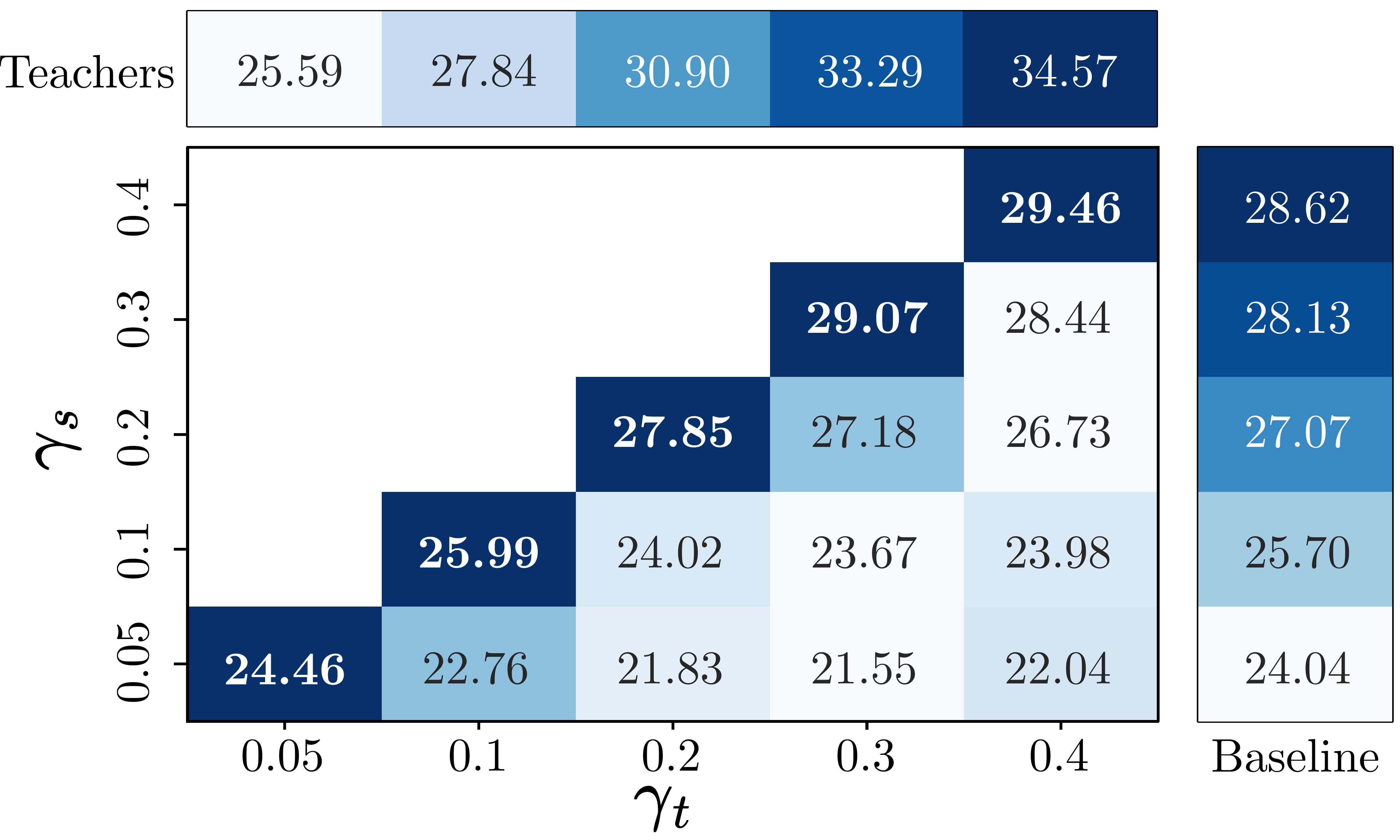}
    \caption{Ablation study using different SPC teachers to optimize the student. Results are reported in PSNR.}
    \label{fig:ABLATION_TEACHERS_SPC}
\end{figure}

In SPC, teachers with real-valued CAs with $\gamma_t \in \{0.05, 0.1, 0.2, 0.3, 0.4\}$ were used to distill students with binary-valued CAs with $\gamma_s \in \{0.05, 0.1, 0.2, 0.3, 0.4\}$ under the condition $\gamma_t \ge \gamma_s$. Figure \ref{fig:ABLATION_TEACHERS_SPC} shows the results, which indicate that the maximum reconstruction performance is obtained when $\gamma_s = \gamma_t$. This is due to the encoder loss function $\mathcal{L}_{\text{ENC}}$, which minimizes the discrepancy between the Gram matrices of the teacher's and student's forward models. When $\gamma_s = \gamma_t$, the structure of the teacher’s and student’s sensing matrices is more closely aligned, making it easier for the student to approximate the teacher's forward model. However, when $\gamma_t > \gamma_s$, the teacher has access to a greater number of effective measurements due to the higher $\gamma_t$. This leads to a Gram matrix for the teacher that represents a richer set of measurement correlations, which the student, constrained by fewer sampling points, struggles to replicate.

\begin{figure}[t!]
    \centering
    \includegraphics[width=0.75\linewidth]{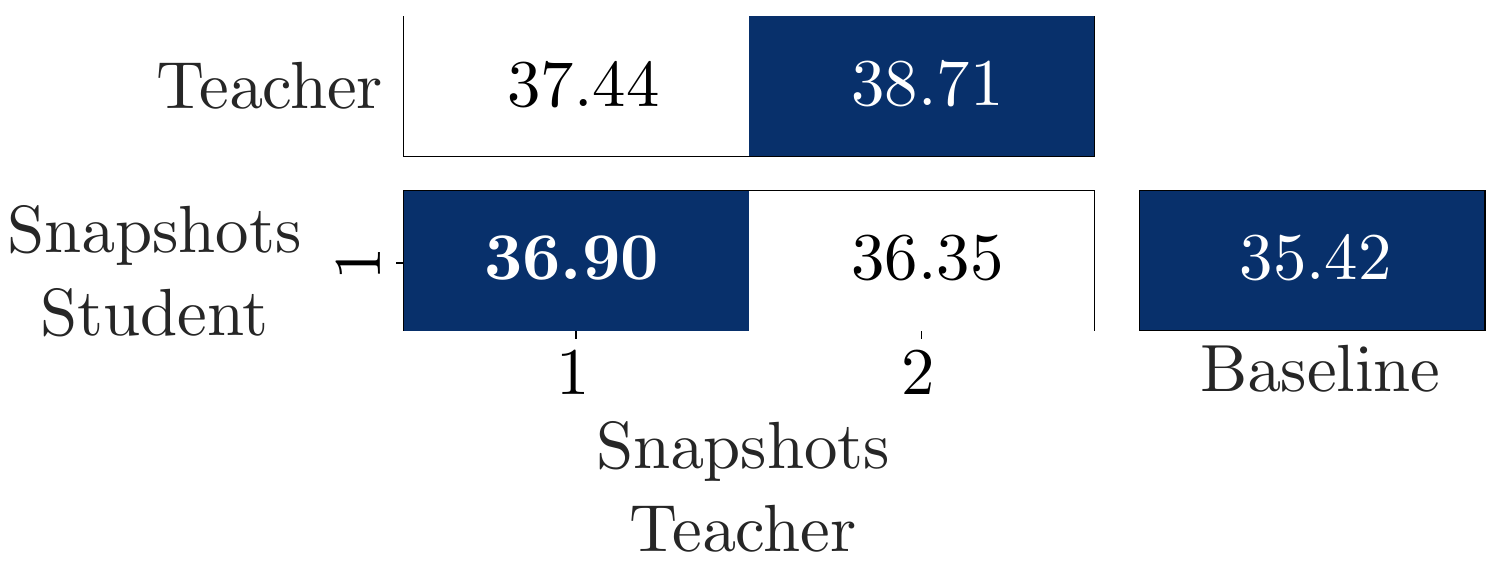}
    \caption{Performance of student models (SD-CASSI with one snapshot and binary-valued CA) distilled from teacher systems configured with one or two snapshots and real-valued CAs. Results are reported in PSNR.}
    \label{fig:TEACHERS_CASSI}
\end{figure}

In the SD-CASSI system, two teacher configurations were evaluated. The first configuration uses a one-snapshot SD-CASSI with a real-valued CA, while the second employs a two-snapshot SD-CASSI with real-valued CAs. Figure \ref{fig:TEACHERS_CASSI} presents the PSNR results obtained by distilling knowledge from these two teacher configurations into a student with a one-snapshot setup and a binary-valued CA. The results reveal a behavior similar to that observed in SPC: the student achieves maximum performance when its number of snapshots matches that of the teacher (one snapshot). This can be attributed to the encoder loss function, $\mathcal{L}_{\text{ENC}}$, which minimizes the discrepancy between the student’s and teacher’s CA Gram matrices. When the number of snapshots is the same for both the student and the teacher, the structure of the teacher’s Gram matrix closely resembles what the student can replicate, facilitating the learning process. However, when the teacher uses more snapshots than the student, the student struggles to replicate the teacher’s Gram matrix, resulting in reduced performance.

\section{Knowledge distillation loss function ablation studies}

In this section, the contributions of the different components of the KD loss function in  are evaluated across the three CI systems. Table \ref{tab:ablation_kd_loss} presents the results obtained using various configurations of the KD loss function. The findings indicate that each loss term contributes individually to improved reconstruction performance compared to the baseline. Notably, the encoder loss function provides the most significant improvement, as it directly aligns the student’s forward model with the teacher’s. Additionally, in almost all cases, combining the two distillation loss terms with the task loss function yields the greatest improvements, highlighting the benefits of jointly distilling both the encoder and decoder components.

\begin{table}[!t]
\centering
\caption{Ablation study on the KD loss function components ($\mathcal{L}_{\text{ENC}}$, $\mathcal{L}_{\text{DEC}}$) across three CI systems. Results are presented in terms of PSNR and SSIM.}

\label{tab:ablation_kd_loss}
\begin{tabular}{ccccc}
\hline
\multicolumn{5}{c}{MRI}\\ \hline
\multicolumn{1}{c|}{$\mathcal{L}_{\text{ENC}}$} &\ding{51}&\ding{51}&\ding{56}&\ding{56}\\
\multicolumn{1}{c|}{$\mathcal{L}_{\text{DEC}}$} &\ding{56}&\ding{51}&\ding{51}&\ding{56}\\
\multicolumn{1}{c|}{PSNR $\uparrow$}           & 41.64 & $\mathbf{42.56}$ & 41.38 & 41.25 \\
\multicolumn{1}{c|}{SSIM $\uparrow$}           & 0.964 & $\mathbf{0.973}$ & 0.961 & 0.960  \\ \hline

\multicolumn{5}{c}{SPC}\\ \hline
\multicolumn{1}{c|}{$\mathcal{L}_{\text{ENC}}$} &\ding{51}&\ding{51}&\ding{56}&\ding{56}\\
\multicolumn{1}{c|}{$\mathcal{L}_{\text{DEC}}$} &\ding{56}&\ding{51}&\ding{51}&\ding{56}\\
\multicolumn{1}{c|}{PSNR $\uparrow$}           & 29.40 &  $\mathbf{29.52}$ & 28.78 &  28.62 \\
\multicolumn{1}{c|}{SSIM $\uparrow$}           & 0.956 &  $\mathbf{0.961}$ & 0.956 &  0.955  \\ \hline

\multicolumn{5}{c}{SD-CASSI}\\ \hline
\multicolumn{1}{c|}{$\mathcal{L}_{\text{ENC}}$} &\ding{51}&\ding{51}&\ding{56}&\ding{56}\\
\multicolumn{1}{c|}{$\mathcal{L}_{\text{DEC}}$} &\ding{56}&\ding{51}&\ding{51}&\ding{56}\\
\multicolumn{1}{c|}{PSNR $\uparrow$} & $\mathbf{36.90}$ &  36.66     &  35.86     & 35.42 \\
\multicolumn{1}{c|}{SSIM $\uparrow$} & $\mathbf{0.958}$ &  0.957    & 0.949 & 0.945 \\ \hline

\end{tabular}
\end{table}

\section{Computational requirements comparison}

In this section, we compare the computational costs of the proposed KD method and the baseline for optimizing the three CI systems. Table \ref{tab:computational_cost} presents the results comparing the proposed KD method with the baseline E2E method, focusing on GPU memory usage and training time for the CI systems: MRI with $AF=4$, SPC with $\gamma=0.4$, and SD-CASSI with one snapshot. These results reveal that while the KD method requires more resources than the baseline E2E method, the increases are relatively modest across all three CI systems. For MRI, the KD approach adds 13 minutes to the training time and requires 1,586 MB more memory. In SPC, the increase is 8 minutes in training time and 82 MB in memory. The largest difference is observed for SD-CASSI, where KD adds 50 minutes and 5,898 MB in memory usage, due the the features map of spectral images being significantly larger than the feature maps of one or two channel images. Despite these increases, the KD proposed approach is justified by the significant improvements in reconstruction performance and encoder design, as demonstrated in previous sections.

Additionally, the inference time remains the same for both the baseline and the proposed method because, after the knowledge transfer process during training, the teacher model is no longer needed. The student model, which has already learned the necessary knowledge from the teacher, performs inference independently. As a result, there are no additional computational costs during inference.

\begin{table}[!t]
\centering
\caption{Comparison of computational cost between the proposed method (KD) and the baseline (E2E) for the three CI systems: MRI with $AF=8$, SPC with $\gamma=0.4$, and SD-CASSI with one snapshot. The metrics include execution time and memory usage.}

\label{tab:computational_cost}
\begin{tabular}{l|cc|cc}
\hline
\multirow{2}{*}{System} & \multicolumn{2}{c|}{Training Time} & \multicolumn{2}{c}{Memory Usage (MB)} \\ \cline{2-5} 
                        & E2E & KD                     & E2E & KD                     \\ \hline
MRI        & 22m 18s & 35m 13s     &   4,084   &  5,670                         \\
SPC        & 12m 12s & 20m 22s     &  1,280    & 1,362                         \\
SD-CASSI   &  2h 13m 20s    &  3h 3m 4s   &  2,405    & 8,303                         \\ \hline
\end{tabular}
\end{table}

\vfill

\end{document}